**OpTiDDM (Optical Tweezers integrating Differential Dynamic Microscopy) maps the spatiotemporal propagation of nonlinear stresses in polymer blends and composites**

Karthik R. Peddireddy, Ryan Clairmont, Philip Neill, Ryan McGorty, and Rae M. Robertson-Anderson*

*Department of Physics and Biophysics, University of San Diego, 5998 Alcala Park, San Diego, CA 92110, United States*

*randerson@sandiego.edu

**Abstract**

How local stresses propagate through polymeric fluids, and, more generally, how macromolecular dynamics give rise to viscoelasticity are open questions vital to wide-ranging scientific and industrial fields. Here, to unambiguously connect polymer dynamics to force response, and map stress propagation in macromolecular materials, we present a powerful approach–Optical Tweezers integrating Differential Dynamic Microscopy (OpTiDMM)–that simultaneously imposes local strains, measures resistive forces, and analyzes the motion of the surrounding polymers. Our measurements with blends of ring and linear polymers (DNA) and their composites with stiff polymers (microtubules) uncover a surprising resonant response, in which affine alignment, superdiffusivity, and elastic memory are maximized when the strain rate is comparable to the entanglement rate. Microtubules suppress this resonance, while substantially increasing elastic force and memory, due to varying degrees to which the polymers buildup, stretch and flow along the strain path, and configurationally dissipate stress. More broadly, the rich multi-scale coupling of mechanics and dynamics afforded by OpTiDDM, empowers its interdisciplinary use to elucidate non-trivial phenomena that sculpt stress propagation dynamics– critical to commercial applications and cell mechanics alike.

**Introduction**

Entangled polymeric fluids, blends and composites–ubiquitous in the natural and commercial world–exhibit complex viscoelastic behavior[1-5], with the characteristics of stress dissipation and storage dictated by the dynamics and relaxation mechanisms of the comprising polymers, inter-polymer interactions and entanglements, and the nature of stress propagation and distribution through the network. The relative dissipation versus storage of induced stress, and the associated spatiotemporal scales, rely on how polymers move, deform, rearrange, stretch, and distribute stress to neighboring polymers.

Despite this critical coupling of polymer dynamics with material stress response and propagation, current techniques to probe entangled polymers, as well as other complex fluids and soft materials, are primarily designed to measuring either force response or macromolecular transport and dynamics. For example, bulk rheology measures the viscoelastic properties and stress response of the bulk fluid in response to macroscopic shearing. On the other hand, typical microrheology measurements track the Brownian fluctuations of individual microscale probes embedded in the material to extract viscoelastic properties using the generalized Stokes-Einstein relation (GSER)[6]. Variations to the GSER approach have employed dynamic light scattering (DLS)[7,8] and differential dynamic microscopy (DDM)[9,10] to extract viscoelastic properties from Brownian fluctuations. These microscale methods require <10 μL and are able to resolve spatial heterogeneities inaccessible to bulk rheology, but the passive nature of the measurements restricts applicability to the linear regime. Moreover, choosing an appropriate probe size and surface treatment for a given system is both non-trivial and essential for correctly interpreting the data[11-13]. Active microrheology methods employing, e.g., magnetic or optical tweezers, are able to apply stress or strain in a local region of the fluid and measure the force the polymers exert to resist the strain[14-16]. For example, using optical tweezers (OpT) microrheology a trapped microscale probe can be pulled through the fluid with precisely controlled rates and distances such that the force response in the nonlinear regime (in which the fluid is pushed far from equilibrium by fast and/or large strains) and at mesoscopic scales (10s of microns) can be measured[17].

To complement these rheological measurement techniques, macromolecular dynamics and transport can be measured by tracking the Brownian fluctuations of single fluorescent-labeled molecules using particle-tracking algorithms[15,18-20]. DLS[7,8] and DDM[9,10,21] can also characterize dynamics and transport by examining ensemble-averaged macromolecular fluctuations. DDM is particularly advantageous for certain systems as it can probe larger spatiotemporal scales than particle-tracking, and can extract information from weaker signals in which single molecules cannot be resolved[9,10,21,22]. The deleterious effects of photobleaching are also comparatively reduced as measurement accuracy does not rely on tracking single polymers over extended time periods. DLS has similar advantages as DDM but does not use microscope images as its input so cannot be coupled to other microscopy techniques such as OpT in a straightforward manner.

In efforts to couple stress response to macromolecular dynamics, opto-rheometers have been used to correlate bulk rheological response with macromolecular deformation, and microfluidic shearing devices have been used to measure polymer stretch and relaxation under extensional flow[23-30]. These methods have shed important light on the dynamics of single polymers under uniform strain and flow, revealing, e.g., shear banding and tumbling of entangled linear DNA under bulk shear[31], and heterogeneous relaxation and swelling of ring DNA under extensional micro-flows due to transient threading by linear chains[32]. However, these techniques require relatively large sample volumes (~$10^2$-$10^3$ μL) which limit their use to study highly entangled biopolymer solutions and other valuable or liable biomaterials which are time and resource intensive to prepare. Moreover, these methods are not equipped to measure scale-dependent, hierarchical phenomena that often emerge in polymer mixtures, or map how stress from a local deformation or disturbance is distributed and propagated through the network.

The complexity of available relaxation mechanisms, transport modes, inter-polymer interactions, and conformational dynamics of entangled polymers–greatly amplified for blends and composites of polymers with different sizes, stiffnesses and topologies[19,32-44]–demands techniques that can unequivocally correlate polymer transport and dynamics to force response and relaxation, and elucidate the spatiotemporal characteristics of stress propagation throughout these systems. For example, in blends of ring and linear polymers, threading of rings by linear chains has been shown to lead to emergent enhanced viscosity and shear-thinning, along with longer relaxation timescales and suppressed terminal regime scaling[35,45-50]. Threaded rings have been shown to both swell and collapse depending on the sizes of the rings and linear chains and undergo more heterogeneous transport than their linear counterparts[51]. At the same time, ring-ring entanglements are weaker and less persistent than their linear counterparts leading to faster relaxation and reduced shear-thinning owing to their reduced ability to retain entanglements and entropically stretch in the shear direction[49,52]. While these diverse phenomena suggest rich and complex modes of spatiotemporal stress propagation and dissipation, how polymers in ring-linear blends dissipate and distribute induced stress to the neighbors remains completely unexplored.

Polymer topology also plays an important role in the force response of composites of flexible and stiff polymers. For example, previous studies on composites of DNA and rigid microtubules (MT) showed that composites comprising linear DNA undergo microscale phase separation and MT flocculation; however, exchanging linear DNA for rings of the same size and concentration promotes DNA-MT mixing and hinders

MT polymerization[43]. These highly distinct architectures give rise to a dramatic non-monotonic dependence of force on MT concentration for composites with linear DNA, with the resistive force first increasing then decreasing as MT concentration is increased, due to increasing MT flocculation that reduces the connectivity and percolation of the network. Conversely, DNA-MT composites formed with ring DNA exhibit much smaller and monotonic increase in force with increasing MT concentration. These findings suggest that DNA-MT composites composed of comparable concentrations of ring and linear DNA may be most effective at maintaining MT connectivity and conferring high strength and resistance to deformation. The polymer dynamics that lead to this hypothesized behavior have yet to be explored.

Because MTs are $10^4\times$ stiffer than DNA ($l_{p,MT} \approx 1$ mm vs $l_{p,DNA} \simeq 50$ nm)[53-55] and ~10× thicker (~25 nm vs 2 nm)[53,56], their response to induced stress, and the spatiotemporal scales over which they can deform to dissipate and distribute stress, are widely different than those for DNA. For example, DNA polymers typically assume random coil configurations in steady-state, as opposed to the extended profile of rigid rod MTs. Further, DNA under shear can entropically stretch, bend, dis-entangle and reorient to dissipate stress, while MTs are much more limited in their dynamical response to imposed stress[36,41,57,58]. Timescales associated with configurational relaxation, e.g., the disengagement time $\tau_D$ over which a polymer reptates out of its entanglement tube, are orders of magnitude longer for MTs compared to DNA[11,44,53,59,60], such that entangled MT solutions respond largely elastically, with minimal viscous dissipation, while entangled DNA displays viscoelasticity with terminal flow behavior in response to strains that are slower than $\tau_D$.

Here, we introduce a powerful experimental technique–Optical Tweezers integrating Differential Dynamic Microscopy (OpTiDDM)–to directly measure the macromolecular deformations and dynamics induced by local linear and nonlinear disturbances and map the associated stress propagation field. Specifically, we apply continuous cycles of nonlinear straining in a local region of a polymer network while simultaneously imaging single fluorescent-labeled polymers (DNA) surrounding the strain and measuring the force the network exerts on the probe. We demonstrate unambiguous coupling between macromolecular dynamics and resistive forces in response to strain, to reveal nontrivial relationships between the stress response, dissipation, macromolecular strain alignment and flow, and propagation of stress and strain. Our measurements with blends of ring and linear DNA and their composites with stiff microtubules demonstrate the power of OpTiDDM to discover unexpected physical phenomena and dissect complex and subtle relationships between various metrics of dynamics and mechanics. For example, we show that while resistive forces increase monotonically with increasing strain rate for all networks, stress propagation dynamics show a unique non-monotonic rate dependence that depends on the structural composition of the network. Moreover, while DNA-MT composites exhibit the strongest elastic-like force response, highly entangled DNA blends exhibit the most pronounced propagation field and scale-dependent dynamics.

**Results and Discussion**

**OpTiDDM (Optical Tweezers integrating Differential Dynamic Microscopy) couples macromolecular dynamics to stress response and propagation**

Because this work is the first introduction and implementation of OpTiDDM (Optical Tweezers integrating Differential Dynamic Microscopy), and a primary aim of the work is to enable other researchers to use OpTiDDM to investigate a wide range of polymeric fluids, macromolecular networks and soft materials, we first describe the critical components and rationale for the technical design, which is based on a force-measuring optical tweezers (OpT) outfitted with an epifluorescence microscope (Fig 1). In our setup, an optically trapped microsphere probe (Fig 1a) can be translated through a sample at precise distances (up to 50 μm) and rates (up to 100s of μm/s) using a piezoelectric mirror to move the trap or a piezoelectric stage to move the sample relative to the trap[17]. We measure the force the sample exerts on the probe to resist the strain using a position-sensing detector (PSD). These features allow us to impart nonlinear and mesoscale strains and measure the resulting local stress response (Fig 1b). We have previously established these OpT microrheology protocols for measuring linear and nonlinear stress response of a range of polymer networks and soft materials[11,33,35,40,41,59-64]. We note that moving the trap while keeping the sample fixed allows for precise imaging of the surrounding sample (that is not moving) over time before, during and following imposed strains, as the sample remains fixed in the field-of-view (FOV) for the duration of the measurement. On the other hand, moving the sample while keeping the trap fixed ensures a constant trapping potential to facilitate accurate force calibration and measurement[17].

To visualize the quiescent Brownian motion and strain-induced non-equilibrium motion of polymers comprising the fluids, we use the fluorescence capability of our OpT-enabled microscope to image fluorescent-labeled DNA tracer molecules embedded in the sample and filling a 78μm×117μm FOV centered on the strain path of the probe (Fig 1c). We record time-series of the moving DNA tracers before, during and after the strain induced by the moving probe.

Our strain program, which we designed to accurately measure polymer deformation and relaxation in response to nonlinear straining and optimize statistical significance and signal-to-noise, consists of cyclically sweeping an optically trapped probe forward and backward through a horizontal distance of $s = 15$ μm at controlled strain rates $\dot{\gamma}$ that span a ~20-fold dynamic range ($\dot{\gamma} = 9.4 - 189$ s$^{-1}$). The maximum and minimum rates are determined by the strength of the optical trap and the requirement of >5 oscillations before noticeable photobleaching occurs, respectively. Each oscillatory shear persists for 50 s, including cessation periods of $\Delta t_R = 3$ s between each sweep to allow the network to relax (Fig 1b), such that each independent measurement includes 6 – 9 strain cycles (depending on $\dot{\gamma}$) (Fig 1). We record the force exerted on the probe at 20 kHz for the full 50 s duration of each cyclic measurement.

To determine the impact of cyclic straining on the polymer dynamics and map the deformation field and propagation of stress imposed by the strain, we perform differential dynamic microscopy (DDM) on spatially-resolved regions of interest (ROI), each (16.6 μm)$^2$, centered horizontally (along $x$) with the strain path and vertically (orthogonal to the strain) at distances of $y = 8$ μm $\simeq s/2$ to $y = 27$ μm $\simeq 2s$ from the strain path (Fig 1c). Optimal characterization of the dynamics requires a high density of labeled molecules to spatially resolve statistically robust dynamics within each ROI, and a lower signal threshold than standard single-particle-tracking (SPT) methods to facilitate using DNA molecules or other macromolecules that are smaller, dimmer, and more susceptible to photobleaching over long acquisition times, as compared to microspheres and other standard probes used in SPT. DDM is optimal for these needs as it can extract dynamics from microscope images that are prohibitively dense or noisy for conventional SPT, by examining image differences in Fourier space to quantify ensemble density fluctuations[9,10]. While implementations of DDM often use microsphere tracers to probe network dynamics, here we use the comprising polymers themselves to directly visualize and report the macromolecular dynamics (Fig 1c). We have previously established and validated the use of DDM for characterizing both passive and active transport of DNA and other biopolymers in crowded and entangled environments[19,20,25,38,39,65,66].

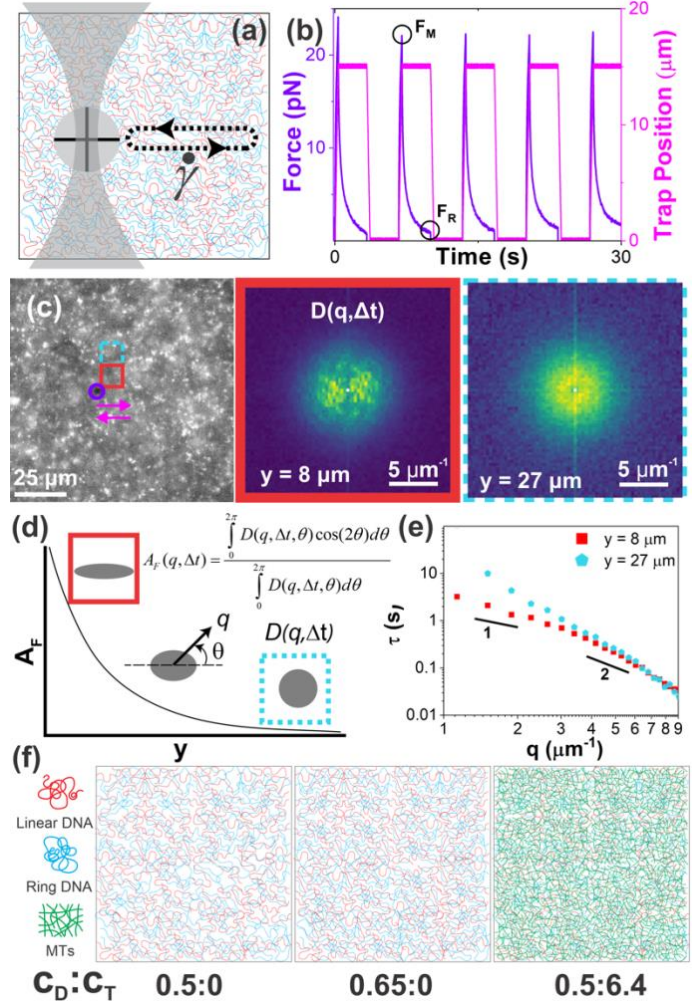

**Fig 1. OpTiDDM integrates optical tweezers microrheology with differential dynamic microscopy to directly couple local stress response to macromolecular dynamics and stress propagation in complex polymeric fluids.**
**(a,b)** Optical tweezers (OpT): **(a)** A microsphere probe (grey circle) embedded in a polymer network is trapped using a focused laser beam, and cyclically swept back and forth through a distance $s = 15$ μm at strain rates of $\dot{\gamma} = 9.4, 19, 42, 90,$ and $189$ s$^{-1}$ for a total of 50 s, including $\Delta t_R = 3$ s cessation periods between each sweep. **(b)** During each measurement, performed at a different $\dot{\gamma}$, we measure the trap position (magenta) and force exerted on the probe (purple), and evaluate the maximum force reached during each sweep $F_M$ and the residual force at the end of each cessation period $F_R$. **(c-e)** Differential Dynamic Microscopy (DDM): **(c)** During each strain, we collect time-series of fluorescent-labeled DNA molecules in the networks. Magenta circle and arrows indicate the probe position and direction of motion during strain. We divide the FOV into (16.6 μm)$^2$ ROIs centered horizontally with the strain path at 10 vertical distances centered at $y = 8$ μm (red border) to $y = 27$ μm (blue-dashed border) from the strain path, and determine the corresponding DDM image structure function $D(\vec{q}, \Delta t)$ as a function of lagtime $\Delta t$ and wavevector $q$. $D(q_x, q_y)$ at $\Delta t = 0.25$ s for boxed-in ROIs show that, near the strain ($y = 8$ μm), $D(\vec{q}, \Delta t)$ is anisotropic and preferentially aligned along the strain direction while far from the strain ($y = 27$ μm) $D(\vec{q}, \Delta t)$ is radially symmetric. **(d)** The alignment factor, $A_F(y)$, schematically portrayed and computed using the equation shown, quantifies the preferential alignment of $D(\vec{q}, \Delta t)$ with the strain path, generally decreasing with increasing $y$ as qualitatively depicted. **(e)** Corresponding DDM decay times $\tau(q)$ determined by fitting $D(q, \Delta t)$ (see Methods) describe the macromolecular dynamics, with diffusive and ballistic dynamics described by $\tau(q) \sim q^{-2}$ and $\tau(q) \sim q^{-1}$ respectively. **(f)** Cartoons of the three polymer networks we use to demonstrate OpTiDDM, defined by the mass concentration (mg/ml) of the ring-linear DNA blend $c_D$ (0.5 or 0.65) and molar concentration (μM) of tubulin dimers $c_T$ (0 or 6.4).

There are several important metrics that DDM can extract from the time-series by computing an image structure function $D(\vec{q}, \Delta t)$ for varying wavevectors $q$ of each pair of images separated by a lag time $\Delta t$ (Fig 1c-e)[9,10,21,67]. The first metric we examine is the alignment factor $A_F$, which we adopt from scattering experiments[68], which assesses the anisotropy of $D(\vec{q}, \Delta t)$ along a given axis (the strain path direction $x$ in our experiments) to determine the degree to which motion is preferred along that direction versus randomly distributed with an isotropic $D(\vec{q}, \Delta t)$, in which case $A_F$ is approximately zero.

The second important metric is the $q$-dependent density fluctuation decay time $\tau(q)$, which describes the type (e.g., ballistic, diffusive, subdiffusive, etc.) and rate (e.g., speed, diffusion coefficient, generalized transport coefficient) of the dynamics. Specifically, we compute $\tau(q)$ by fitting the radially averaged image structure function to $D(q, \Delta t) = A(q)[1 - f(q, \Delta t)] + B(q)$, where $A(q)$ is the amplitude, $B(q)$ the background, and $f(q, \Delta t) = e^{-(\Delta t/\tau(q))^\delta}$ the intermediate scattering function that describes the dynamics[9,10]. $\tau(q)$ can often be described by a power-law, $\tau(q) \simeq (Kq)^{-\beta}$, where $\beta=1$, $\beta=2$, and $\beta>2$ corresponds to ballistic motion, diffusion and subdiffusion, and $K$ is the associated speed, diffusion coefficient and generalized transport coefficient. For reference, the relationship between $\beta$ and the more widely known anomalous scaling exponent $\alpha$ in the relationship between the mean-squared displacement and lag time, i.e., $MSD \sim t^\alpha$, is $\beta = 2/\alpha$.

Finally, we examine the stretching exponent $\delta$ in the function we use to fit $f(q, \Delta t)$. Diffusive (Brownian) motion in dilute and isotropic thermal systems exhibit $\delta=1$ scaling[9,69]. However, superdiffusive or ballistic-like motion, as seen in actively driven systems, often manifest $\delta>1$[70-72], and subdiffusive dynamics that crowded and heterogeneous systems exhibit are typically better fit to $\delta<1$[20,22,66,69,73].

To demonstrate the efficacy, sensitivity, and wide applicability of OpTiDDM, we perform measurements on entangled solutions of blended ring and linear DNA molecules with equal contour lengths of $L = 115$ kbp $\simeq 38$ μm, as well as their composite with microtubules (MT) (Fig 1f). In all three systems we investigate, the ring:linear DNA ratio is fixed to 1:1, a ratio which has been shown to give rise to enhanced elastic strength and shear-thinning compared to the corresponding single-component systems[35,74,75], as well as maximal propensity for threading of rings by linear chains[35,42]. We examine blend concentrations of 0.5 mg/ml (~$3c_e$ where $c_e$ is the critical entanglement concentration[76]) and 0.65 mg/ml (~$4c_e$), as well as a composite of the 0.5 mg/ml DNA blend and MTs polymerized from 6.4 μM tubulin. The MT concentration was chosen such that the corresponding entanglement tube diameter $d_T$ is comparable to that of the 0.65 mg/ml DNA solution (see Methods) such that we can isolate the dependence of polymer stiffness on the phenomena we discover. Throughout the paper, we refer to these three systems by their corresponding DNA mass concentration $c_D$ (in mg/ml) and tubulin molar concentration $c_T$ (in μM), i.e., $c_D : c_T = 0.5:0$, $0.65:0$, and $0.5:6.4$. Importantly, these systems are only subtly different from one another, with <25% DNA concentration differences, minimal changes in entanglement spacing and constant length and topologies of the DNA. In this way, we are able to demonstrate the sensitivity of OpTiDDM to measure variations in the dynamics and stress response of networks that are minimally distinct.

**Alignment of entangled DNA with the strain exhibits non-monotonic dependence on strain rate**

We first examine the anisotropy of $D(\vec{q}, \Delta t)$ to determine the degree to which DNA motion surrounding the local strain deviates from isotropic thermal fluctuations and preferentially aligns with the strain path, i.e. affine deformation. As shown in Figure 1c, close to the strain path ($y = 8$ μm $\simeq s/2$), the image structure function is clearly anisotropic, with two lobes centered to the left and right of $q = (0,0)$, indicating preferred motion along the $x$-axis (parallel to the strain path). Conversely, $D(\vec{q}, \Delta t)$ far from the strain path ($y = 27$ μm $\simeq 2s$) is radially symmetric and centered at $q = (0,0)$.

To quantify the strain alignment, we compute the alignment factor $A_F(y)$ as described above (Fig 1d). As shown in Fig 2a, for an intermediate strain rate of $\dot{\gamma} = 42$ s$^{-1}$, $A_F$ decreases to zero as the distance from the strain path $y$ increases for all three systems. The high-concentration DNA blend (0.65:0) exhibits the strongest alignment closest to the strain, but, interestingly, the decay of $A_F$ with increasing $y$ is markedly stronger than for the other two systems (0.5:0 and 0.5:6.4). We next turn to $A_F(y)$ curves for the slowest (9.4 s$^{-1}$) and fastest (189 s$^{-1}$) strain rates (Fig 2b), which are ~4.5-fold lower and higher than $\dot{\gamma} = 42$ s$^{-1}$ examined in Fig 2a. In general, the faster $\dot{\gamma}$ (189 s$^{-1}$) results in a higher degree of alignment closest to the strain path and steeper decay with $y$ than the slow rate (9.4 s$^{-1}$). However, surprisingly, $A_F(y)$ for both rates are lower than those for $\dot{\gamma} = 42$ s$^{-1}$ for all three systems. In other words, strain alignment does not scale monotonically with $\dot{\gamma}$ as one might expect. This non-monotonicity, clearly shown in Fig 2c, is strongest for the 0.65:0 network and weakest for the DNA-MT composite (0.5:6.4), such that the DNA-MT composite has the strongest alignment among the three systems at low and high $\dot{\gamma}$, while the 0.65:0 blend is the system with the strongest alignment at $\dot{\gamma} = 42$ s$^{-1}$.

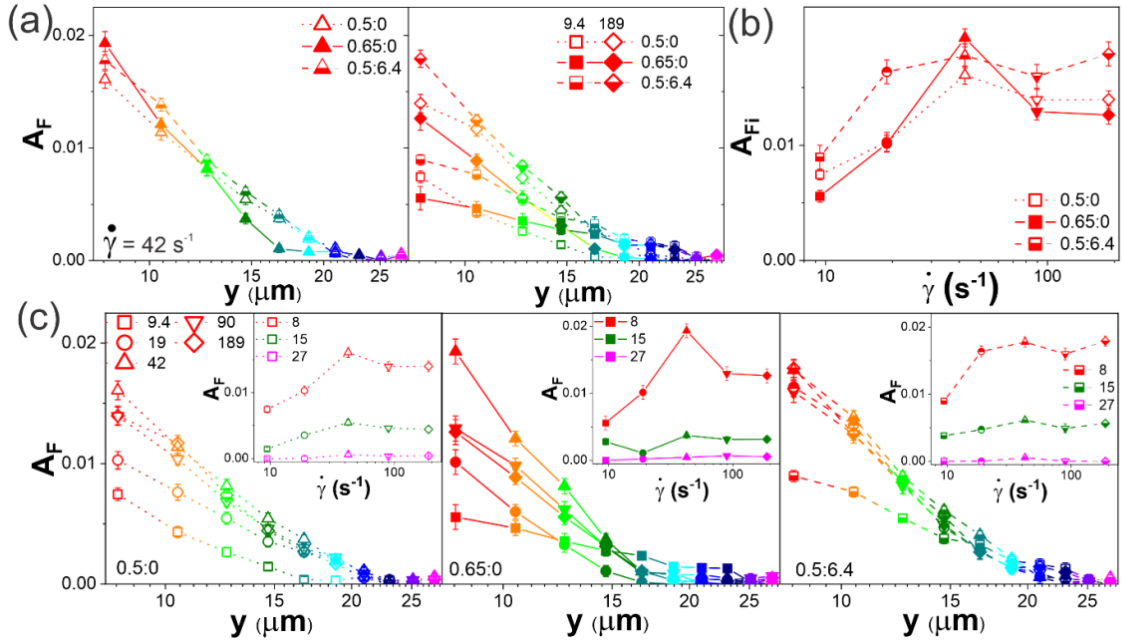

**Figure 2. Cyclic strain induces affine alignment of DNA with a non-monotonic rate dependence that is suppressed by microtubules.** **(a)** Alignment factor $A_F$ versus distance $y$ (increasing from warm to cool colors) for a $\dot{\gamma} = 42$ s$^{-1}$ strain (left) and ~4.5-fold slower ($\dot{\gamma} = 9.4$ s$^{-1}$, squares) and faster ($\dot{\gamma} = 189$ s$^{-1}$, diamonds) strains (right) applied to solutions with $c_D$:$c_T$ = 0.5:0 (open symbols), 0.65:0 (closed symbols), and 0.5:6.4 (half-filled symbols). **(b)** $A_{F,i}$, the alignment factor for the closest ROI ($y = 8$ μm), versus $\dot{\gamma}$ for all three systems (listed in legend) reveals a non-monotonic $\dot{\gamma}$ dependence that is most pronounced for 0.65:0 and weakest when MTs are present (0.5:6.4). **(c)** $A_F(y)$ for strain rates (s$^{-1}$) of $\dot{\gamma}$=9.4 (squares), 19 (circles), 42 (upright triangles), 90 (inverted triangles), and 189 (diamonds) for (left) 0.5:0, (middle) 0.65:0 and (right) 0.5:6.4 systems. Insets show the $\dot{\gamma}$ dependence of $A_F$ near ($y = 8$ μm), far ($y = 27$ μm) and at intermediate distances ($y = 15$ μm) from the strain. Both DNA blends (left, middle) show nonmonotonicity that is stronger closer to the strain, while the DNA-MT composite (right) exhibits minimal $\dot{\gamma}$ dependence for $\dot{\gamma} > 9.4$ s$^{-1}$.

To more closely evaluate the $\dot{\gamma}$ dependence of the small-$y$ alignment as well as the lengthscale overwhich strain-induced alignment decays to zero, we plot $A_F(y)$ for all rates for each system separately (Fig 2c). $A_F(y)$ for the 0.5:0 blend displays non-monotonic $\dot{\gamma}$ dependence for all distances from the strain path and the lengthscale over which $A_F$ decays follows a similar trend. Both the strength and propagation of alignment are maximized at $\dot{\gamma}$ =42 s$^{-1}$. The higher concentration DNA blend exhibits the same non-monotonic dependence near the strain path but for $y > 15$ μm ($\simeq s$), the slowest strain rate surprisingly exhibits the strongest alignment while $A_F$ for $\dot{\gamma}$ =42 s$^{-1}$ is undetectable. This strong alignment propagation for the lowest strain rate is also evident in the DNA-MT composite, suggesting that strong connectivity is required for this phenomenon. The slow rate minimizes potential strain-induced dis-entangling, de-threading, shear-thinning or spatial heterogeneity–all of which would compromise the connectivity of the networks. Conversely, the strong decay at this slow rate for the 0.5:0 system, which has fewer entanglements than the other two systems, indicates that the polymers can more easily deform and move to dissipate induced stress, thereby reducing the propagation of stress to surrounding polymers.

Further, different than the pure DNA blends, the DNA-MT composite exhibits minimal $\dot{\gamma}$ dependence of $A_F$ for $\dot{\gamma} > 9$ s$^{-1}$. This insensitivity to $\dot{\gamma}$, a hallmark of elasticity, suggests that rigid MTs suppress the viscous dissipation enabled by flexible DNA molecules deforming, stretching and reorienting in response to the strain. The inability of MTs to undergo entropic stretching available to DNA random coils likely also plays a role in this phenomenon. We expect entropic stretching of DNA from random coils to more extended affinely-aligned conformations to be stronger with more entanglements (i.e., stronger for 0.65:0 than 0.5:0), as it requires that each polymer is pulled in the direction of the strain while also being substantially pulled in the opposite direction by the entangling polymers that are trailing or are further from the strain. This effect, seen in Fig 2c, would become stronger as $\dot{\gamma}$ increases until $\dot{\gamma}$ becomes faster than the entanglement rate $v_e = \tau_e^{-1}$ where $\tau_e$ is the entanglement time, i.e., the time it takes for each entanglement segment to 'feel' its tube confinement from surrounding entangling polymers[1,60,77,78]. The predicted entanglement times for the 0.65:0 and 0.5:0 blends are $\tau_{e,0.65} \simeq 15.7$ ms and $\tau_{e,0.50} \simeq 26.6$ ms, respectively (see Methods), corresponding to rates of $v_{e,0.65} = \tau_{e,0.65}^{-1} \simeq 64$ s$^{-1}$ and $v_{e,0.50} = \tau_{e,0.50}^{-1} \simeq 38$ s$^{-1}$. These rates are remarkably similar to $\dot{\gamma}= 42$ s$^{-1}$ in which we see maximal affine alignment. For faster strain rates relative to the entanglement rate, i.e., $\dot{\gamma} > v_e$ (90 s$^{-1}$, 189 s$^{-1}$), the polymers do not have time to feel the tube constraints that facilitate stretching, resulting in less affine alignment. At the same time, for slow strain rates, i.e., $\dot{\gamma} < v_e$ (9.4 s$^{-1}$, 19 s$^{-1}$), the entangled polymers have ample time to relax imposed stress on the timescale of the strain, also resulting in less deformation and flow near the strain site. Only for $\dot{\gamma} \approx v_e$ are the polymers confined by entanglements but unable to relax imposed stress, manifesting as the pronounced alignment.

**DNA exhibits strain-induced superdiffusivity and quiescent subdiffusion at lengthscales dictated by the entanglement density**

We next seek to elucidate the DNA dynamics that give rise to the deformation profiles shown in Fig 2, by evaluating the fluctuation decay times $\tau(q)$ for varying distances $y$ from the strain path as described above (Fig 1e). We first focus on the $\dot{\gamma}$ =42 s$^{-1}$ data for the 0.65:0 blend, as it exhibits the most pronounced strain alignment (Fig 2). As shown in Fig 3a,b, $\tau(q)$ curves for all $y$ values collapse to a single curve for $q \gtrsim 4$ μm$^{-1}$, with approximately diffusive scaling of $\beta_2 \approx 2$ suggestive of diffusive dynamics and minimal strain impact. However, for $q \lesssim 4$ μm$^{-1}$, there is a stark monotonic decrease in the scaling exponent as we move closer to the strain site (cool to warm colors indicate $y$ =27 μm to $y = 8$ μm). To better show this effect we scale $\tau$ by $q^{\beta_2}$, where we determine $\beta_2$ by fitting $\tau(q)$ for $4 \lesssim q \lesssim 7$ μm$^{-1}$ (Fig 3b). Positive scaling of $\tau q^{\beta_2}$ curves near the strain (warm colors, small $y$), are indicative of superdiffusive or ballistic dynamics ($\beta_1 \rightarrow 1$), whereas negative scaling, seen far from the strain site (cool colors, large $y$), indicates subdiffusive dynamics ($\beta_1 > 2$). The $y$-dependence of $\tau(q)$ scaling suggests that the strain alignment is a result of DNA moving preferentially (nearly ballistically) in the direction of the strain until $y \approx 15$ µm, at which point $\beta_1 > \beta_2$ and DNA motion is dominated by tube confinement that results in subdiffusive dynamics. The distance at which this crossover occurs is similar to that at which $A_F$ becomes negligibly small.

Further, the wavevector $q_c$ at which scaling crosses over from $y$-dependent $\beta_1$ scaling, indicative of superdiffusion or subdiffusion, to $y$-independent diffusive scaling $\beta_2$, corresponds to a lengthscale $\lambda_c = 2\pi/q_c \simeq 1.6$ µm, remarkably close to the predicted tube diameter of the DNA solution $d_T \simeq 1.5$ µm (see Methods). At lengthscales smaller than the tube diameter, the effects of entanglements are not felt by the DNA segments so there is no perceived confinement (thus no subdiffusion), nor is there perceived 'pulling' from entanglements that are moving along with the strain (thus no directed motion or flow).

These general trends in dynamics are also seen for the less entangled DNA blend (0.5:0) and the DNA-MT composite (0.5:6.4) (Fig 2c,d). However, the wave vector at which $\beta_1$ values diverge for different $y$ distances is slightly smaller ($q_c \approx 3.4$ µm$^{-1}$), as is the corresponding spread in $\beta_1$ values. In other words, the disturbance has less of an effect on these solutions compared to the 0.65:0 blend. The crossover lengthscale in both cases is $\lambda_c = 2\pi/q_c \approx 1.8$ µm, close the the tube diameter of $d_T \simeq 1.7$ µm for the 0.5:0 DNA blend (see Methods), corroborating our interpretation for the 0.65:0 crossover. We also note that the magnitude of $\tau(q)$ is higher for 0.65:0 than the two other solutions across nearly all $q$ values, indicating that the dynamics, whether dominated by Brownian diffusion, confined subdiffusion or flow, are slower.

The similarity in $\lambda_c$ for the 0.5:0 and 0.5:6.4 solutions implies that the microtubules have little effect on the lengthscale at which confinement effects are felt. The mesh size $\xi$ of the microtubules in the 0.5:6.4 composites, comparable to the corresponding tube diameter, is $\xi \simeq 0.89 c_T^{-1/2} \simeq 1.1$ µm (here $c_T$ is in mg/ml). This lengthscale corresponds to $q \approx 6$ µm$^{-1}$, close to the largest $q$ value we can accurately measure. As such, any effects from the confinement by the microtubules would be apparent over the entire $q$ range. Surprisingly, comparing Fig 3c and 3d, we see minimal effect of MTs, with the most apparent difference between 0.5:0 and 0.5:6.4 being the reduced effect of strain at large lengthscales, seen as lower $\tau q^{\beta_2}$ values for small $q$ and large $y$.

To better visualize and quantify these complex effects of strain, entanglements and network rigidity on DNA dynamics, we examine the low-$q$ ($\lambda \gtrsim d_T$) and high-$q$ ($\lambda \lesssim d_T$) scaling exponents, $\beta_1$ (Fig 3e) and $\beta_2$ (Fig 3f) respectively, as functions of $y$ for all three systems. $\beta_2$ values show minimal $y$ dependence for all systems, as reflected in Fig 2a-c, indicating that below entanglement lengthscales, the induced strain does not affect DNA dynamics. We also point out that all scaling exponents are subdiffusive with an average value of $<\beta_2> \simeq 2.7$. The DNA-MT composite exhibits the least subdiffusion (lowest $\beta_2$ values) for nearly all distances, similar to what has been reported previously for actin-microtubule composites in which rigid confinement by stiff microtubules allowed for more free diffusion at small lengthscales, compared to more dynamic and flexible actin filaments in which subdiffusion arose from coupling of the confined polymers to the slow rearrangement of their entanglements[19,20,38,66]. In support of this interpretation is the 0.5:0 data which surprisingly exhibits the largest deviation from free diffusion. Fewer entanglements per chain allows for faster rearrangement of entanglements compared to 0.65:0 blends, such that more rearrangement can occur on the timescale of tube confinement leading to stronger subdiffusive scaling.

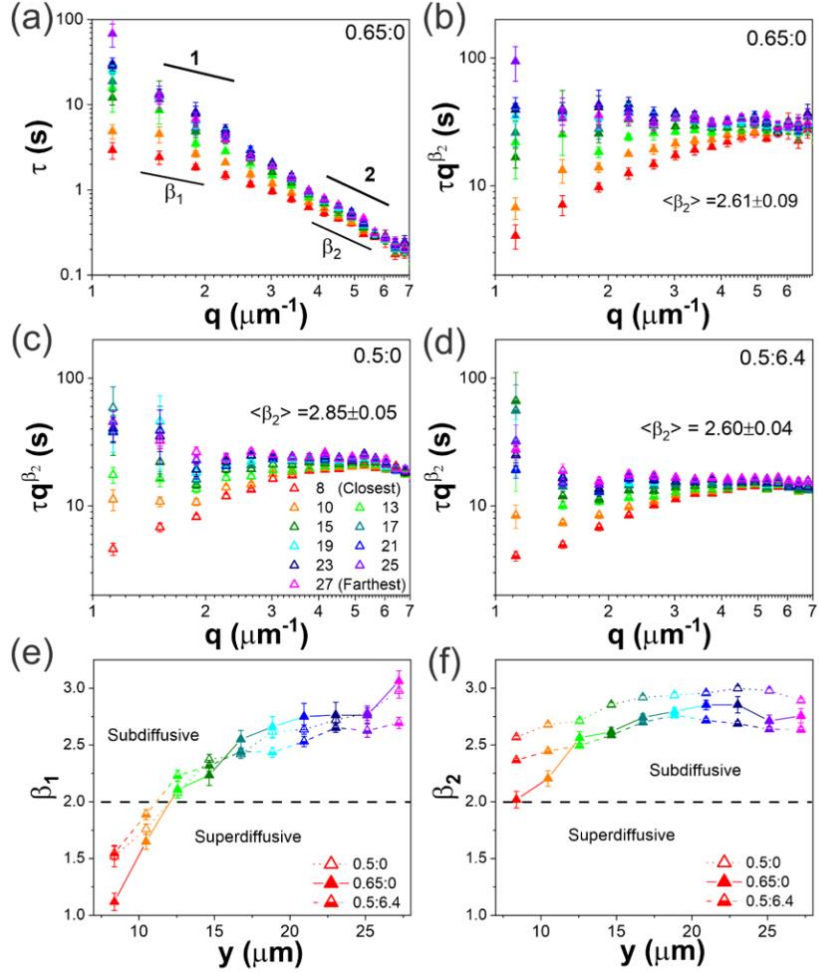

**Figure 3. DDM decay times for $\dot{\gamma}$ = 42 s⁻¹ reveal universal biphasic DNA dynamics that range from ballistic to subdiffusive depending on the distance from the strain.** All data is for $\dot{\gamma}$ = 42 s⁻¹ that exhibits the most pronounced affine alignment (see Fig 2). **(a)** DDM decay times $\tau(q)$ versus wavevector $q$ for the 0.65:0 system evaluated at varying distances $y$, represented as warm to cool colors, for $y$ = 8 to 27 μm (see legend). Closer to the strain, $\tau(q)$ curves exhibit two distinct scaling regimes described by $\tau(q) \sim q^{-\beta_1}$ and $\tau(q) \sim q^{-\beta_2}$ for small and large $q$ values, respectively. Scaling exponents for diffusive ($\beta = 2$) and ballistic ($\beta = 1$) dynamics are also shown. $\tau(q)$ curves for $q \gtrsim 4$ μm⁻¹ are largely independent of $y$, while they diverge for $q \lesssim 4$ μm⁻¹. The crossover wavevector $q_c \simeq 4$ μm⁻¹ corresponds to a lengthscale $\lambda_c = 2\pi/q_c \approx 1.6$ μm, comparable to the tube diameter $d_T \approx 1.5$ μm for 0.65:0. **(b-d)** $\tau(q)$ normalized by the corresponding large-$q$ scaling $q^{-\beta_2}$ versus $q$ for **(b)** 0.65:0, **(c)** 0.5:0, and **(d)** 0.5:6.4 subject to $\dot{\gamma}$ = 42 s⁻¹ straining clearly shows that all systems exhibit biphasic scaling behavior, with strong $y$ dependence for $q \lesssim q_c$, where $\beta_1$ varies from $< \beta_2$ (positive slopes) to $> \beta_2$ (negative slopes) as $y$ increases, compared to $q \gtrsim q_c$ where minimal $y$ dependence (data collapse) is seen. The crossover from $\beta_1$ to $\beta_2$ scaling occurs at a lower $q_c$ value of ~3.3 μm⁻¹ for 0.5:0 and 0.5:6.4 compared to 0.65:0, corresponding $\lambda_c = 2\pi/q_c \approx 1.9$ μm, comparable to $d_T \approx 1.7$ μm for the 0.5:0 system. **(e)** $\beta_1$ versus $y$ determined from fitting $\tau(q < q_c)$ curves for all 3 systems plotted in (b-d). Dashed line indicates normal Brownian diffusion ($\beta_1 = 2$) while $\beta_1 = 1$ and $\beta_1 > 2$ indicate ballistic and subdiffusive dynamics, respectively. 0.65:0 (closed symbols) exhibits the most ballistic-like motion near the strain site and most extreme subdiffusion far from the strain, whereas 0.5:6.4 (half-filled symbols) exhibits the weakest $y$ dependence and the least deviation from normal diffusion. **(f)** $\beta_2$ versus $y$ determined from fitting $\tau(q > q_c)$. $\beta_2$ for all systems exhibit minimal $y$ dependence and weakly subdiffusive dynamics.

Another likely contribution to the subdiffusive dynamics we measure is the threading of ring DNA by linear DNA and microtubules, which we have previously shown to be pervasive in equimolar ring-linear DNA blends[42]. Threading confines the motion of ring polymers to within their radius of gyration $R_{G,R}$ for lateral motion along a threading chain. Threading confinement, which persists on much longer timescales than entanglements[42,51,79,80], leads to more pronounced subdiffusion than that from entanglements alone at lengthscales $\lambda > R_G \approx 0.6$ μm ($q \lesssim 10$ μm$^{-1}$)[81]. As such, threading likely plays a principal role in the large-$y$ subdiffusion we measure across our entire $q$ range ($\lesssim 7$ μm$^{-1}$).

Unlike $\beta_2$, the larger lengthscale ($\lambda \gtrsim d_T$) scaling exponent $\beta_1$ exhibits significant dependence on distance from the strain path, increasing from $\beta_1 \approx 1$ (ballistic) to $\beta_1 \approx 3$ (subdiffusive with $\alpha \approx 0.6$) for the 0.65:0 blend as $y$ increases from 8 to 27 $\mu m$. The enhanced subdiffusion seen at large distances ($y \gtrsim 20$ μm) compared to $\beta_2$ (i.e., $\beta_1 > \beta_2$) likely arises from the onset of entanglement effects that only contribute significantly to the dynamics at lengthscales beyond $d_T$. This effect is most pronounced for the 0.65:0 blend and least pronounced for the 0.5:6.4 composite. At the opposite extreme, close to the strain path ($y \lesssim 15$ μm) the 0.65:0 blend exhibits the most pronounced directional flow, with near-ballistic scaling, while the DNA-MT composite displays the smallest deviation from diffusive dynamics ($\beta_1 > 1.5$).

Previous rheological studies of ring-linear blends have shown that threading of rings by linear chains facilitates their alignment with shear flow, manifesting as shear thinning[35,49,52,82], suggesting that threading likely plays an important role in DNA alignment and superdiffusivity near the strain site. The more pronounced alignment and flow for 0.65:0 blends compared to the other cases indicates that threading is most pervasive in the 0.65:0 blend. This assumption is trivial when considering the 0.5:0 blend, which has fewer total rings and linear chains than the 0.65:0 blend, thus fewer threading events. Reduced threading in DNA-MT composites, as compared to the 0.65:0 blend, is less obvious as previous studies have reported that DNA can indeed become threaded by MTs[19,38]. Threading in these previously reported cases was not the principal mechanism dictating the observed dynamics, in part because of the relatively large diameter of microtubules (~25 nm) compared to the conformational size of the DNA rings (~$R_G \approx 550$ nm), as well as their rigidity and length that suppress thermal reconfiguration into threaded conformations and limits the number of free ends available for rings to thread compared to thinner and more flexible DNA. Finally, while threading of rings by MTs may occur in the 0.5:6.4 composite, there are 23% fewer rings in this system than the 0.65:0 system (~250 vs ~325 μg/mL), so correspondingly fewer threading events can occur.

Finally, the reduced $y$-dependence of $\beta_1$ for the DNA-MT composite may be an indication of the rigid scaffold that the MTs provide to the DNA that hinders network rearrangement and stretching. DNA in different rigid cages formed by the entangled MT network diffuse largely independently of each other as the cages are able to absorb imposed stresses with much less deformation thereby propagating stress to neighboring regions of the network with less dissipation and restructuring. This effect would manifest as less ballistic-like flow and stretching near the strain site, coupled with reduced subdiffusion far from the strain (as described above), such that $\beta_1$ and $\beta_2$ are both closer to 2 than the blends that lack MTs.

**Deformation dynamics of DNA blends display emergent resonant coupling with strain rate that is suppressed by microtubules**

The results described above suggest that the presence of MTs should result in minimal $\dot{\gamma}$ dependence of dynamics, as compared to DNA blends, as the MTs are essentially rigid rods on the timescales of the strains, i.e, their relaxation rates are slower than the strain rates we use. As such, MTs are unable to deform in response to the strain, regardless of $\dot{\gamma}$, such that the dynamics of the DNA confined in the microtubule cages will likewise show minimal $\dot{\gamma}$ dependence. To verify this interpretation, in Fig 4 we examine the $\dot{\gamma}$ dependence of the dynamics and stress propagation presented in Fig 3.

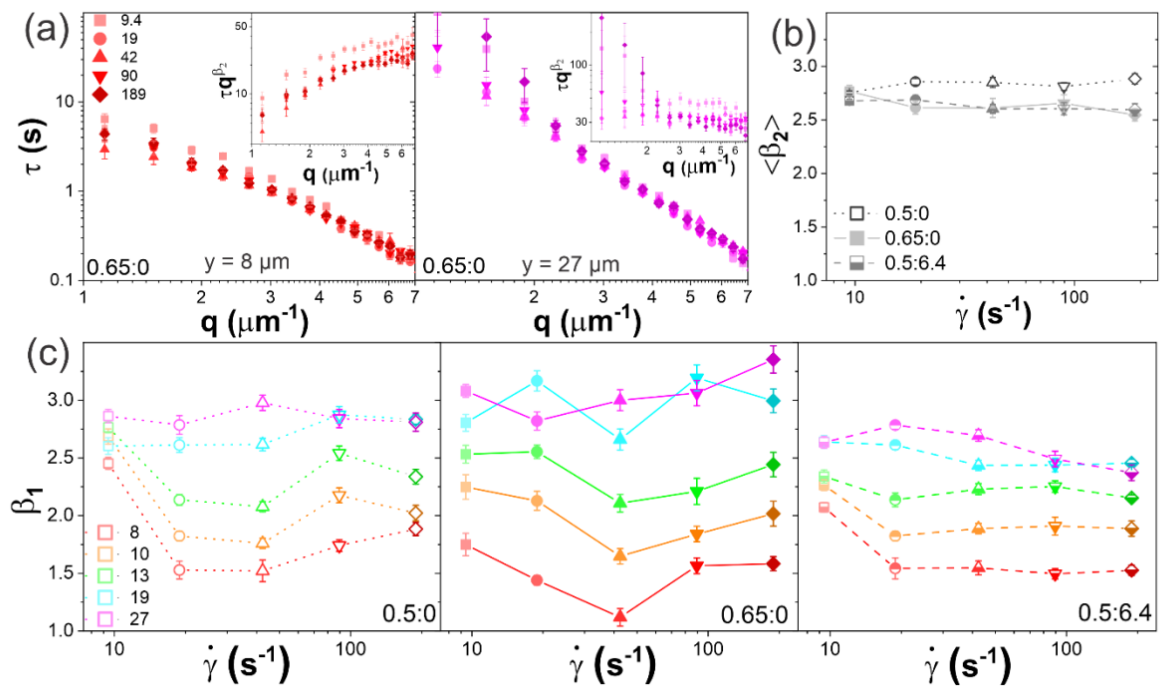

**Figure 4. DNA blends under strain exhibit flow-like dynamics that are maximized at 'resonant' strain rates and decay to subdiffusive dynamics with increasing distance from the strain.** (a) $\tau(q)$ versus wavevector $q$ for the 0.65:0 system evaluated for the (left) nearest and (right) farthest ROIs (i.e., $y = 8$ μm (red) and 27 μm (purple)) subject to strains of $\dot{\gamma}$=9.4 s$^{-1}$ (squares), 19 s$^{-1}$ (circles), 42 s$^{-1}$ (upright triangles), 90 s$^{-1}$ (inverted triangles), and 189 s$^{-1}$ (diamonds), with color gradients from light to dark indicating increasing $\dot{\gamma}$. Insets: Scaled decay times $\tau q^{\beta_2}$ versus $q$ showing the $\dot{\gamma}$-dependent biphasic behavior with (left) near ballistic motion near the strain site for $q \lesssim q_c$ versus (right) subdiffusion far from the strain for $q \lesssim q_c$. (b) $\tau(q)$ scaling exponent $\beta_2$ determined from fits to $\tau(q > q_c)$ and averaged over all $y$ values, show minimal dependence on $\dot{\gamma}$ or system. Error bars indicate the standard error across all $y$ values. (c) $\tau(q)$ scaling exponents $\beta_1$ determined from fits to $\tau(q < q_c)$ versus $\dot{\gamma}$ for varying $y$ distances from the strain, listed in μm in the legend and color-coded from warm to cool colors for increasing $y$, for (left) 0.5:0, (middle) 0.65:0 and (right) 0.5:6.4.

As shown, the $\dot{\gamma}$ dependence both near and far from the strain is weaker than the $y$-dependence shown in Fig 3 (Fig 4a). However, the same two-phase scaling behavior, with $\beta_1$ crossing over to $\beta_2$ at $\lambda_c \approx d_T$ is preserved. Further, similar to the $y$ dependence shown in Fig 3e,f, $\beta_2$ is largely insensitive to $\dot{\gamma}$ or the details of the system and indicates subdiffusion with an average value of $< \beta_2 > \simeq 2.7$ (corresponding to $< \alpha > \simeq 0.74$) (Fig 4c). Distance-dependent $\beta_1$ values, plotted as a function of $\dot{\gamma}$ for each system in Fig 4c, show minimal $\dot{\gamma}$ dependence for the DNA-MT composite (0.5:6.4) and for DNA blends far from the strain site ($y > 15\ \mu m$). Further, the average $\beta_1$ values for $y > 15\ \mu m$, $< \beta_1 >_{\dot{\gamma}} \simeq 2.8$, 3.0 and 2.5, for the 0.5:0, 0.65:0 and 0.5:6.4 systems, respectively, show that subdiffusion is strongest for 0.65:0 in which threading is expected to be most pervasive, and weakest for 0.5:6.4 owing to the rigid caging from the MT scaffold.

However, for $y < 15$ μm, $\beta_1$ for both DNA blends (0.65:0, 0.5:0) displays a similar non-monotonic $\dot{\gamma}$ dependence as we observe for $A_F$, with the most pronounced superdiffusive flow emerging at $\dot{\gamma} = 42$ s$^{-1}$, and the $\dot{\gamma}$ dependence and deviation from diffusive scaling being stronger for 0.65:0 compared to 0.5:0. This observation–indicating maximal entropic stretching and flow at a strain rate of $\dot{\gamma} = 42$ s$^{-1}$ that is 'resonant' with the entanglement rates $\upsilon_{e,0.65}$ and $\upsilon_{e,0.50}$–is in line with our interpretation in the previous section. Namely, when $\dot{\gamma}$ is substantially faster than $\upsilon_e$, the polymers do not feel entanglements that promote flow alignment and stretching, thereby reducing superdiffusivity. At the same time, for $\dot{\gamma} < \upsilon_e$ the polymers have time to relax on the timescale of the strain, resulting in less deformation and affine flow (i.e., superdiffusivity) near the strain site.

This pronounced 'resonant' behavior–in which the system response is maximized when the strain rate is comparable to the intrinsic system relaxation rate–is suppressed by microtubules which, due to their rigidity, have slower relaxation timescales. We can estimate a lower bound for the fastest relaxation time for entangled microtubules based on previous studies of actin-microtubule composites with similar microtubule concentrations[36,41]. These studies reported a value of $\tau_e \approx 60$ ms, corresponding to $\upsilon_e \approx 17$ s$^{-1}$, which is slower than all but our slowest strain rate ($\dot{\gamma} = 9.4$ s$^{-1}$). As such, we expect there to be less alignment (smaller $A_F$) and reduced flow ($\beta_1$ closer to 2) in response to the $\dot{\gamma} = 9.4$ s$^{-1}$ strain compared to the higher rates, as we see in Fig 2c and Fig 4c. At this rate the microtubules are able to partially relax and distribute imposed stress on the timescale of the strain. We do not see a subsequent rise in $\beta_1$ or drop in $A_F$ beyond the 'resonant' rate ($\dot{\gamma} \approx \upsilon_e$) in this system, because the resonant response itself is weaker in the presence of microtubules compared to the 0.65:0 blend. This effect, reflected in the smaller maximum $A_F$ and larger maximum $\beta_1$ in the 0.5:6.4 composite compared to the 0.65:0 blend (Figs 2c and 3e, respectively), is an indicator of the rigidity of the microtubules which reduce their ability to dynamically couple with the strain.

**Threading of DNA rings in entangled ring-linear blends leads to confinement-dominated deformation dynamics across broad spatiotemporal scales**

To verify our interpretations above and further elucidate the effects of confinement, rigidity and strain-induced transport on the polymer dynamics and stress propagation, we examine the stretching exponent $\delta$ in the exponential function we use to fit each intermediate scattering function, $f(q, \Delta t) = e^{-(\Delta t/\tau(q))^\delta}$, as we describe in Methods (Fig 5). Recall that confinement effects often result in $\delta < 1$[20,22,66,69,73] while systems that display active transport typically display $\delta > 1$[70-72]. As shown in Fig 5a-c, when subject to the nominal resonant strain of $\dot{\gamma} = 42$ s$^{-1}$, the stretching exponents for all systems decrease to $\delta \approx 0.5$ as $q$ increases, unlike typical steady-state systems in which $\delta$ is largely independent of $q$.[9,69] Closest to the strain path ($y = 8$ μm), all systems exhibit $\delta > 1$ for small $q$ values, but there is a substantial reduction in low-$q$ stretching exponents as we move further from the strain ($y \rightarrow 27$ μm). In other words, the effect of the strain, which we expect to manifest as $\delta > 1$ consistent with active transport or flow, becomes weaker as we move away from the strain site, similar to our Fig 2 results.

Moreover, the $y$ dependence of $\delta$ is strongest for the 0.65:0 blend (Fig 5b), dropping ~2-fold at low $q$ values and displaying minimal $q$ dependence at large $y$ distances. The 0.5:0 and 0.5:6.4 systems exhibit comparatively weaker $y$ dependence and more $q$ dependence at large $y$ (Fig 5a,c). This effect is likely another indication of increased threading probability in 0.65:0 compared to the other systems. We previously showed that steady-state actin-microtubule composites exhibited $\delta$ values of ~0.5 to 0.7 with the lowest and highest values arising in composites that are highly crosslinked and entangled (no crosslinks), respectively[66]. The fact that our measured $\delta$ values are closer to those reported for crosslinked systems, which have extremely slow relaxation timescales compared to entangled systems, suggests that threaded rings, which also relax much more slowly than entangled rings, play a principal role in the dynamics, in particular for the 0.65:0 system which has the highest concentration of rings.

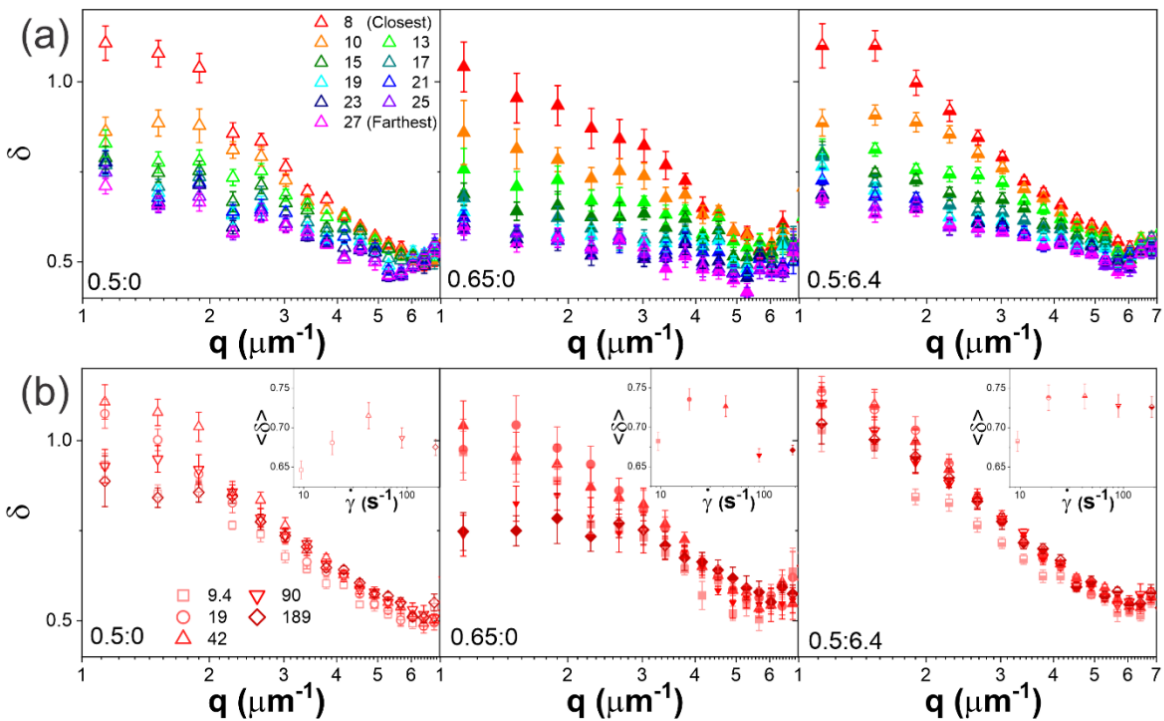

**Figure 5. DDM stretching exponents $\delta$ indicate DNA dynamics similar to active transport near the strain and confined motion far from the strain.** (a) DDM stretching exponents $\delta$ versus wave vector $q$ for varying distances from the strain, listed in μm and color-coded from warm to cool colors for increasing $y$, for the 0.5:0 (left, open symbols), 0.65:0 (middle, filled symbols) and 0.5:6.4 (right, half-filled symbols) systems subject to $\dot{\gamma} = 42$ s$^{-1}$ straining. 0.65:0 exhibits the largest $y$ dependence and lowest values of $\delta$. (b) $\delta(q)$ for $y = 8$ μm (red, closest to the strain) and varying strain rates $\dot{\gamma}$, listed in s$^{-1}$ and delineated by different symbols and a color gradient from light to dark for increasing $\dot{\gamma}$ for 0.5:0 (left, open symbols), 0.65:0 (middle, filled symbols) and 0.5:6.4 (right, half-filled symbols). Insets showing the corresponding $<\delta>$, averaged over $q$, versus strain rate highlight the non-monotonic $\dot{\gamma}$ dependence for the DNA-only systems (left, middle). The presence of MTs quenches this non-monotonicity and generally increases $\delta(q)$.

Similar to the increased $y$ dependence of the stretching exponent for the 0.65:0 blend compared to the other two systems, we observe a similar increased $\dot{\gamma}$ dependence, as shown in Fig 5b. The values of the stretching exponents are also generally lower for the 0.65:0 blend, indicative of stronger confinement. Finally, the signature resonant behavior is also evident in the two DNA blends, with the $\dot{\gamma} = 42$ s$^{-1}$ strain inducing the highest $\delta$ values, consistent with the most directed transport or flow compared to the other strain rates. Conversely, the DNA-MT composite displays minimal $\dot{\gamma}$ dependence and generally higher $\delta$ values, suggestive of reduced subdiffusivity, as we see in Figs 3e and 4c.

**Microtubules enhance force response and suppress shear-thinning in DNA blends**

To directly couple the strain propagation and deformation dynamics to the force response of the networks, we measure the force the polymers exert on the moving probe during the same nonlinear oscillatory shear that we use for the measurements and analyses in Figs 2-5. Specifically, we measure the nonlinear force response during repeated cycles of sweeping the bead back and forth through $s = 15$ μm for a total of 50 s with a $\Delta t_R = 3$ s cessation period between each sweep (Fig 6).

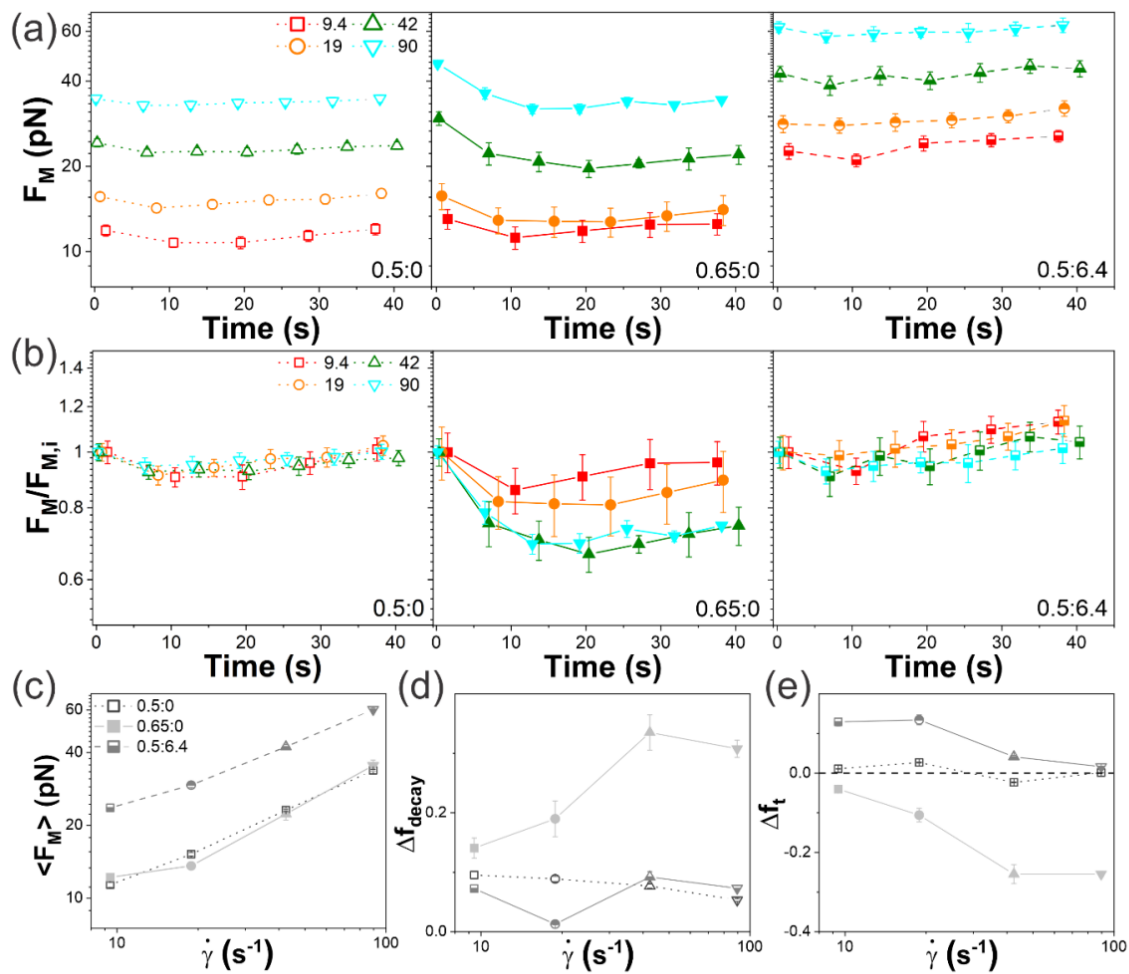

**Figure 6. The force response from cyclic straining monotonically increases with strain rate and is strongest in the presence of microtubules.** (a) Maximum force $F_M$ reached during each cycle of repeated nonlinear straining (see Fig 1) versus measurement time for $\dot{\gamma} = 9.4$ s$^{-1}$ (red), 19 s$^{-1}$ (orange), 42 s$^{-1}$ (olive), and 90 s$^{-1}$ (cyan) are shown for 0.5:0 (left, open symbols), 0.65:0 (middle, filled symbols) and 0.5:6.4 (right, half-filled symbols) systems. (b) Data shown in (a), normalized by the corresponding initial force $F_{M,i}$, shows that $F_M$ for 0.65:0 (middle) decreases with subsequent cycling in a $\dot{\gamma}$-dependent manner while $F_M$ increases for 0.5:6.4 (right). (c) $F_M$ averaged over all cycles, (d) fractional decay in force during cycling, $\Delta f_{decay} = (F_{M,i} - F_{M,min})/F_{M,i}$, and (e) fractional increase in force over the course of cycling, $\Delta f_t = (F_{M,final} - F_{M,i})/F_{M,i}$ versus $\dot{\gamma}$ for 0.5:0 (open), 0.65:0 (filled), and 0.5:6.4 (half-filled). Note that the force generally increases over the course of cycling for 0.5:0 and 0.5:6.4, suggesting polymer buildup at the strain edges, while the force decreases for 0.65:0, as expected for shear-thinning and flow alignment. All error bars are standard error over 15 trials.

We first evaluate the maximum force $F_M$ reached in each sweep for all three systems at four strain rates $\dot{\gamma}$ = 9.4 – 90 s$^{-1}$ (Fig 6a). Interestingly, all systems display a monotonic increase in force $F_M$ with increasing strain rate $\dot{\gamma}$, despite the non-monotonic trends in the deformation dynamics (Figs 2,4). Further, the DNA-MT composite displays higher $F_M$ values and more pronounced $\dot{\gamma}$ dependence compared to the DNA-only blends (Fig 6a,c), despite the minimal $\dot{\gamma}$ dependence seen in the corresponding deformation dynamics (Fig 2-5). These results suggest that the rigidity and slow relaxation timescales of the MTs limit their ability to rearrange and move along with the strain, instead exerting a strong resistive force in response to the deformation. Conversely, the flexible DNA molecules that can entropically stretch, bend and rearrange, can more easily dissipate the imposed stress by configurationally deforming, stretching and flowing.

While the initial force values (measured for the first 2-3 cycles) are higher for the 0.65:0 DNA blend than 0.5:0 for all strain rates, $F_M$ decays to values comparable to those of the 0.5:0 blend by $t_r \approx 20$ s. Importantly, the fractional drop in force $\Delta f_{decay}$ from the starting value $F_{M,i}$ is greatest for the 'resonant' rate $\dot{\gamma} = 42$ s$^{-1}$ (in which flow and alignment are maximized), with $\Delta f_{decay} = (F_{M,i} - F_{M,min})/F_{M,i} \approx 0.25$ (Fig 6d). Taken together, these results indicate shear-thinning, whereby the viscosity of the solution is reduced in response to strain due to flow alignment. Indeed, previous studies have shown that equal-mass ring-linear blends exhibit enhanced shear-thinning compared to their pure linear or ring counterparts, that is facilitated by ring threading[35,75]. The initial decay in $F_M$ for the 0.65:0 blend further suggests that the longest relaxation timescale has not been reached in the early cycles. Indeed, the predicted disengagement time for linear DNA at 0.65 mg/ml is $\tau_D \approx 23$ s, remarkably close to $t_r \approx 20$ s in which $F_M$ reaches a nearly constant time-independent value. For times longer than $\tau_D$, the DNA can relax its stress equally, regardless of time, such that a steady-state is reached.

For the less entangled DNA system (0.5:0), $F_M$ values remain nearly constant for all cycles, dropping by <10% in the first 2 cycles, followed by a slight increase that is most pronounced for the slowest strain rate, that likewise exhibits the weakest alignment and flow. To understand the subsequent rise in force following the initial decay, which can also be seen for the slower strain rates of the 0.65:0 blend, we turn to our DNA-MT composite results. Rather than thinning, which is minimal for 0.5:6.4, we instead observe a modest increase in force over the course of cycling, reaching $\Delta f_t = (F_{M,final} - F_{M,i})/F_{M,i} \approx 0.12$ for the slowest strain rate, in which the effect is most pronounced (Fig 6e). This increase (Fig 6b,e), is indicative of polymer buildup at the edges of the strain, otherwise thought of as osmotic compression[62], which has previously been evidenced in entangled DNA and actin systems[37,62,64]. This increase is most apparent at the slowest strain rate, which also shows the least amount of flow and alignment, suggesting that the DNA polymers cluster at the leading edge of the moving probe, increasing the local entanglement density and forming a wake in the strain path. Threading events, which are most probable in the 0.65:0 blend, likely reduce this clustering by facilitating flow alignment and enabling de-threading of rings by linear chains, which, in turn, suppress polymer buildup at the leading edges[50,82-84]. Reduced threading probability in the 0.5:0 and 0.5:6.4 systems leads to weaker thinning and more polymer buildup, which is enhanced by the presence of microtubules which cage the DNA and further suppress network relaxation.

**Interplay between strain-induced flow alignment and osmotic compression dictates force dissipation and elastic memory**

To shed further light on the coupling between the viscoelasticity, force response and deformation of the entangled polymers, we next evaluate the force relaxation during each $\Delta t_R = 3$ s cessation period between cycles. Fig 7 shows the residual force $F_R$ maintained at the end of each cessation period as a function of cycling time. Unlike the maximum resistive force $F_M$ reached in each cycle, the residual force $F_R$ is largely independent of $\dot{\gamma}$ for 0.5:0 and 0.5:6.4 and only weakly decreases with increasing $\dot{\gamma}$ for 0.65:0. Moreover, $F_R$ for the 0.5:0 and 0.5:6.4 systems both increase monotonically with time (i.e., cycle number), whereas the 0.65:0 system exhibits non-monotonic time-dependence with the same minimum at $t_r \approx 20$ s (Fig 7a).

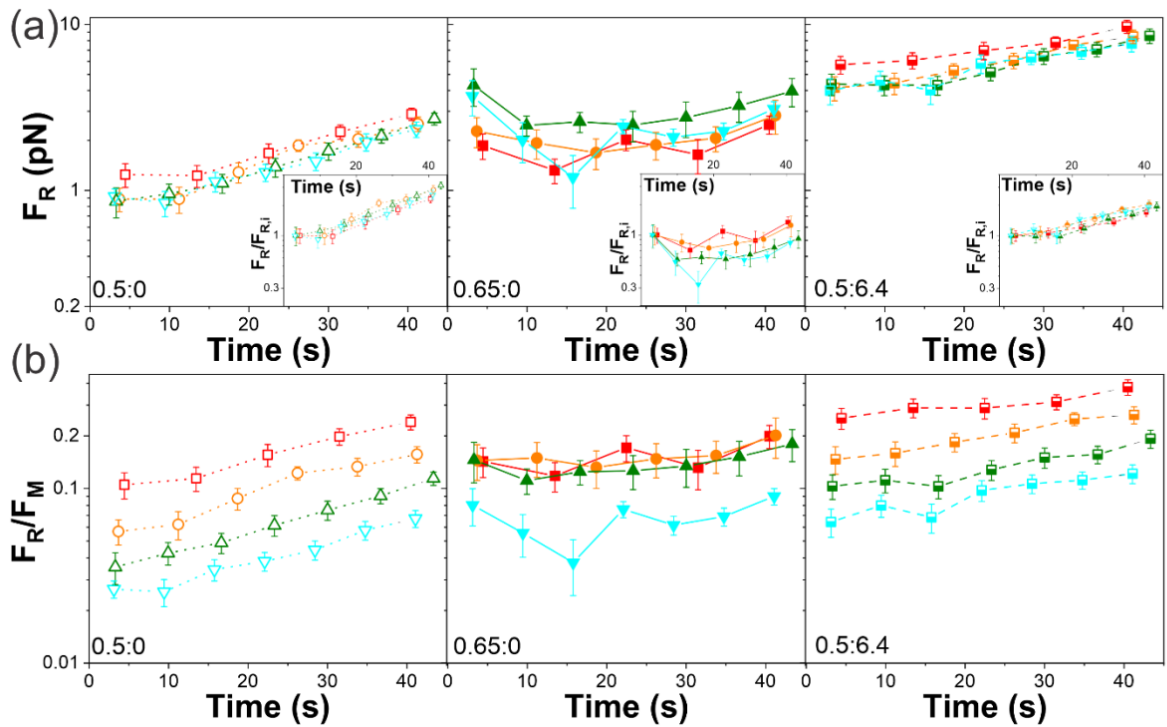

**Figure 7. The residual force maintained following the cessation period is surprisingly $\dot{\gamma}$-independent with markedly different time dependence for the 0.65:0 blend.** (a) Residual force $F_R$ maintained at the end of each $\Delta t_R = 3$ s cessation period versus measurement time during repeated straining (see Fig 1) at $\dot{\gamma}$ = 9.4 s$^{-1}$ (red), 19 s$^{-1}$ (orange), 42 s$^{-1}$ (olive), and 90 s$^{-1}$ (cyan) for 0.5:0 (left, open symbols), 0.65:0 (middle, filled symbols) and 0.5:6.4 (right, half-filled symbols). Insets show $F_R$ normalized by the corresponding initial value $F_{R,i}$. (b) Data shown in (a), normalized by the corresponding maximum force $F_M$, shows that the fractional force maintained, an indicator of elastic memory, increases with time and $\dot{\gamma}$ for 0.5:0 and 0.5:6.4, while it is nearly constant for the 0.65:0 system, indicating that different mechanisms are driving the force response and relaxation of 0.65:0 compared to 0.5:0 and 0.5:6.4.

The fractional force that is sustained during $\Delta t_R$, which we define as $F_R/F_M$, is a measure of the relative elastic memory versus dissipation due to polymer relaxation and rearrangement. The DNA-MT composite exhibits the highest $F_R/F_M$ values for all strain rates, as we may expect given that the corresponding resistive force is highest and the deformation and flow is lower than both DNA-only systems. Likewise, the 0.5:0 blend exhibits the lowest residual force $F_R/F_M$ for all $\dot{\gamma}$. Interestingly, for both of these systems, 0.5:0 and 0.5:6.4, $F_R/F_M$ decreases with increasing $\dot{\gamma}$ and increases with increasing time (subsequent cycling). Conversely, the 0.65:0 system displays minimal dependence on $\dot{\gamma}$ or cycle.

The lack of $\dot{\gamma}$ or time dependence for 0.65:0 is further evidence of $\dot{\gamma}$-dependent thinning and flow alignment, such that as $\dot{\gamma}$ increases, $F_M$ does not increase to a comparable degree because the polymers are more preferentially aligned with the strain path. This affine deformation and entropic stretching limits the extent to which polymers buildup at the edges of the strain path, as described above. On the other hand, the increase in elastic memory with cycling in the 0.5:0 and 0.5:6.4 systems is a clear indicator of continued build-up of DNA at the edges of the strain, which is more pronounced in the absence of microtubules (0.5:0). This effect likely arises from the DNA in these two systems have fewer entanglements to restrict their motion, compared to the 0.65:0 system, so they can more readily be pushed by the probe to the strain edge.

Osmotic compression or densification of DNA at the strain edges increases the local entanglement density which would, in turn, result in higher resistive forces that continue to increase with each cycle as more polymers diffuse into the strain wake and then get pushed to the leading edge[64]. This process is clearly dependent on the timescale of the strain. Slower strains would provide more time for the polymers to rearrange and untangle with neighbors to stay at the leading edge of the probe, whereas faster strain rates would result in more polymers slipping off of the moving probe as they are pulled back by entangling polymers. This effect results in $F_R/F_M$ decreasing as $\dot{\gamma}$ increases, as we see for 0.5:0 and 0.5:6.4.

**OpTiDDM couples strain-induced stress and elastic memory to macromolecular motion and affine alignment in response to local disturbances**

To directly couple the macromolecular dynamics we measure with DDM to the local force response we measure using OpT, we evaluate the relationship between key metrics that each measurement produces: the maximum force reached in the initial strain cycle $F_{M,i}$, the residual force at the end of the first cessation period $F_{R,i}$, the alignment factor closest to the strain site $A_{F,i}$, and the low-$q$ $\tau(q)$ scaling exponent $\beta_1$ closest to the strain site ($\beta_{1,i}$). In Fig 8 we directly compare each of these metrics for all strain rates and systems, resulting in six unique plots: $(A_{F,i} \text{ vs } F_{M,i})$, $(\beta_{1,i} \text{ vs } F_{M,i})$, $(F_{R,i} \text{ vs } F_{M,i})$, $(\beta_{1,i} \text{ vs } A_{F,i})$, $(F_{R,i} \text{ vs } A_{F,i})$, and $(F_{R,i} \text{ vs } \beta_{1,i})$. The parameters listed along the top-left to bottom-right diagonal correspond to the $x$ and $y$ axes of each plot, which are color-coded accordingly.

The plots along the diagonal compare two parameters from the same measurement technique, characterizing either force (i.e., $F_{R,i}$ vs $F_{M,i}$, corner plot) or dynamics (i.e., $\beta_{1,i}$ vs $A_{F,i}$, inner plot). These plots corroborate what we have already described in the preceding sections, which we summarize as follows. We interpret the alignment factor as arising from entropic stretching and flow along the strain direction, with increasing $A_{F,i}$ values indicating more alignment. Within this interpretation, increasing $A_{F,i}$ should correspond with decreasing scaling exponent $\beta_{1,i}$, which decreases from $\beta_{1,i} > 2$ for subdiffusive motion to $\beta_{1,i} = 1$ for ballistic motion. This relationship is generally what we see in the orange-bordered plot in Fig 8, in which higher/lower $\beta_{1,i}$ values pair with lower/higher $A_{F,i}$ values. Moreover, we see the least variation in values for different strain rates for the 0.5:6.4 system, whereas we see the largest variation and lowest/highest values of $\beta_{1,i}/A_{F,i}$.

Our OpT measurements show that, in general, the maximum force $F_{M,i}$ increases with increasing $\dot{\gamma}$, while the residual force $F_{R,i}$ is largely independent of $\dot{\gamma}$. Moreover, the 0.5:6.4 system exhibits the largest force response (highest $F_{M,i}$) and least dissipation (highest $F_{R,i}$), while 0.5:0 exhibits the lowest $F_{M,i}$ and $F_{R,i}$. This trend is reflected in the green-bordered plot which shows that ($F_{M,i}$, $F_{R,i}$) values are generally largest for the 0.5:6.4 system, which we attribute to enhanced elasticity from the rigid MTs, and smallest for 0.5:0 as this system has the fewest entanglements and can thus most easily move and rearrange to dissipate strain-induced force. For both 0.5:0 and 0.5:6.4, $F_{M,i}$ increases monotonically with $\dot{\gamma}$ while $F_{R,i}$ is nearly independent of $\dot{\gamma}$. Conversely, the 0.65:0 system, which also shows monotonic increase of $F_{M,i}$ with $\dot{\gamma}$, exhibits a ~2-fold increase in $F_{R,i}$ as $\dot{\gamma}$ increases to the resonant rate $\dot{\gamma} = 42 \text{ s}^{-1}$, reaching a value larger than the corresponding value for the 0.5:6.4 composite, followed by a modest decrease. Notably, for $\dot{\gamma} < 42 \text{ s}^{-1}$ the residual force for 0.65:0 is comparable to the less entangled DNA system (0.5:0) whereas for $\dot{\gamma} \geq 42 \text{ s}^{-1}$, $F_{R,i}$ for 0.65:0 closely matches that of the DNA-MT composite (0.5:6). These data demonstrate a shift in the 0.65:0 data from largely dissipative, similar to the minimally-entangled 0.5:0 system to more elastic-like, similar to the 0.5:6 system, suggestive of the onset of strong entanglements and constraints for strain rates comparable to the entanglement rate, $\dot{\gamma} \approx \nu_e$. The subsequent drop in $F_{R,i}$ for $\dot{\gamma} > 42 \text{ s}^{-1}$ is further indication of shear-thinning and forced de-threading, as described above.

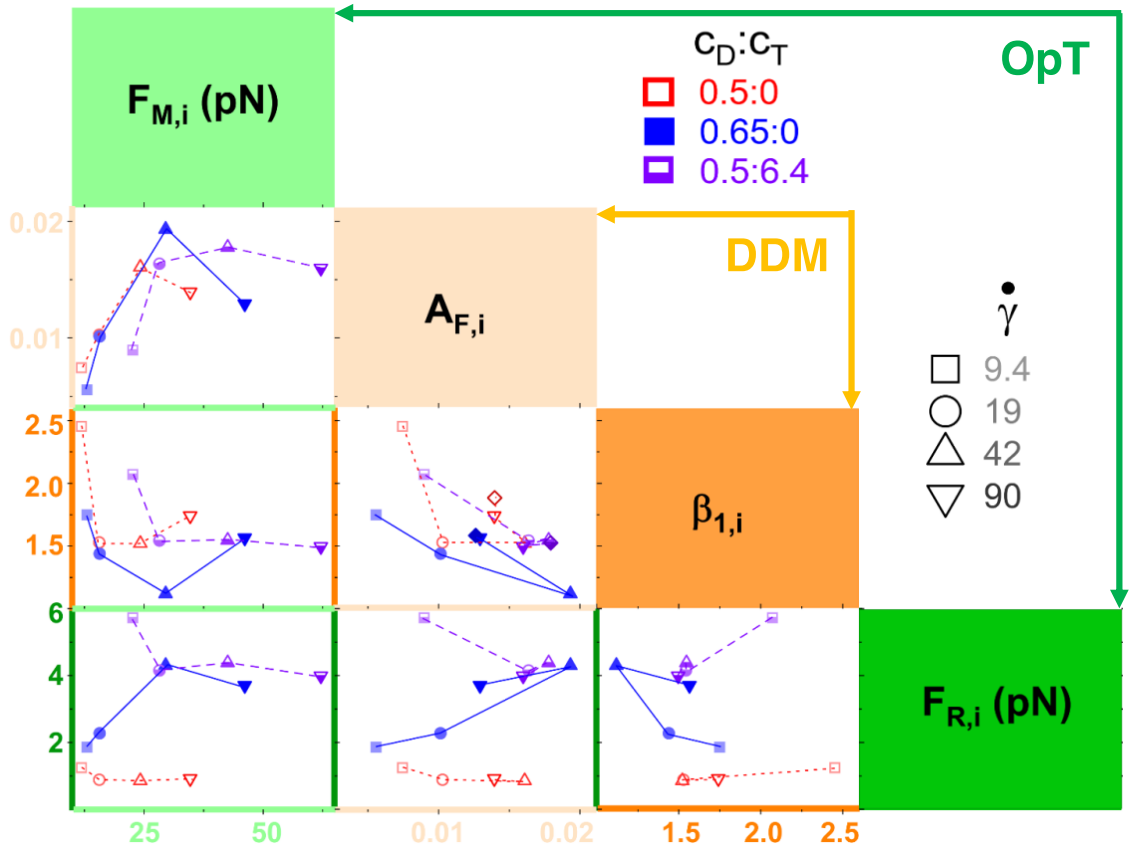

**Figure 8. OpTiDDM maps the local stress response of polymeric materials to the underlying macromolecular dynamics and deformations.** Force-dynamics phase map comparing key metrics from OpT and DDM. $F_{M,i}$ (dark green, top left) and $F_{R,i}$ (light green, bottom right) are the maximum force $F_M$ and residual force $F_R$ for the first strain cycle, measured via OpT. $A_{F,i}$ (dark orange, middle left) and $\beta_{1,i}$ (light orange, middle right) are the alignment factor $A_F$ and $q < q_c$ scaling exponent $\beta_1$ for the closest ROI ($y = 8$ μm), determined via DDM. The resulting six unique metric pairs plotted – $(A_{F,i}, F_{M,i})$, $(\beta_{1,i}, F_{M,i})$, $(F_{R,i}, F_{M,i})$, $(\beta_{1,i}, A_{F,i})$, $(F_{R,i}, A_{F,i})$, $(F_{R,i}, \beta_{1,i})$ – are denoted by the axes colors. In all plots the data shown is for 0.5:0 (red, open), 0.65:0 (blue, filled), and 0.5:6.4 (purple, half-filled) subject to strains of $\dot{\gamma} = 9.4 \text{ s}^{-1}$ (squares, light shades), 19 s$^{-1}$ (circles, medium-light), 42 s$^{-1}$ (triangles, medium-dark) and 90 s$^{-1}$ (inverted triangles, dark). Panels along the diagonal compare metrics from the same method, with OpT ($F_{M,i}$ and $F_{R,i}$) in the corner panel and DDM ($\beta_{1,i}, A_{F,i}$) in the middle panel. The off-diagonal plots directly compare an OpT-measured metric ($F_{M,i}$ or $F_{R,i}$) to a DDM-determined metric ($\beta_{1,i}$ or $A_{F,i}$).

Having established our chosen metrics as good markers for the dynamics and mechanics we reveal in Figs 2-7, we next evaluate the plots that compare dynamics ($A_{F,i}$ and $\beta_{1,i}$) and mechanics ($F_{M,i}$ and $F_{R,i}$). Comparing the alignment factor $A_{F,i}$ and maximum force $F_{M,i}$ (top left plot) reveals that for slow strain rates, $A_{F,i}$ generally increases with increasing $F_{M,i}$ until $\dot{\gamma} = 42$ s$^{-1}$ followed by a decrease. The non-monotonic dependence of $A_{F,i}$ on $F_{M,i}$ is strongest for the 0.65:0 system and weakest for 0.5:6.4, demonstrating maximal affine deformation and flow alignment for the 0.65:0 system. Also notable is the fact that while $A_{F,i}$ values for 0.5:6.4 are comparable to those for the 0.5:0 system, the corresponding $F_{M,i}$ for 0.5:6.4 is substantially higher, indicating that the rigid microtubules require substantially higher forces for similar deformation.

We see similar relations between $\beta_{1,i}$ and $F_{M,i}$ (middle left plot), with $\beta_{1,i}$ generally decreasing with increasing $F_{M,i}$ until $\dot{\gamma} = 42$ s$^{-1}$ after which $\beta_{1,i}$ increases with $F_{M,i}$ for the DNA-only blends (0.5:0 and 0.65:0) while $\beta_{1,i}$ remains constant for the DNA-MT composite (0.5:6.4). Further, the 0.65:0 system displays the most extreme non-monotonicity in $\beta_{1,i}$ with increasing $F_{M,i}$ as well as the lowest $\beta_{1,i}$ values, indicating the most directed, ballistic-like motion and entropic stretching. This flow alignment results in shear-thinning, i.e., reduced increase in $F_{M,i}$ with increasing $\dot{\gamma}$. We also point out that the 0.5:0 system exhibits the lowest $F_{M,i}$ and largest $\beta_{1,i}$ values of all 3 systems, consistent with minimal flow alignment and a high degree of dissipation from thermal rearrangement and osmotic compression of DNA.

How the varying degrees of alignment ($A_{F,i}$) and directed motion ($\beta_{1,i}$) impact the elastic retention of force following the strain ($F_{R,i}$) can be seen in the bottom plots of Fig 8 showing $A_{F,i}$ (column 2) and $\beta_{1,i}$ (columns 3) vs $F_{R,i}$. Similar to their dependence on $F_{M,i}$, $A_{F,i}$ and $\beta_{1,i}$ exhibit mirror dependences on $F_{R,i}$, with all systems displaying non-monotonic $\dot{\gamma}$ dependence, with maxima and minima in $A_{F,i}$ and $\beta_{1,i}$, respectively, occurring at $\dot{\gamma} = 42$ s$^{-1}$. Further, the 0.65:0 system displays the largest range in $A_{F,i}(\dot{\gamma})$ and $\beta_{1,i}(\dot{\gamma})$ values, indicating the largest dynamic range and the strongest resonant coupling of relaxation rates and strain rates. Moreover, the 0.65:0 system switches from exhibiting behavior similar to the 0.5:0 blend to that of the 0.5:6 composite. The latter result once again demonstrates a switching from dissipative (small $F_{R,i}$) to elastic-like (large $F_{R,i}$) dynamics for the 0.65:0 blend at $\dot{\gamma} \approx \nu_e$, while the other two systems display either dissipative (0.5:0) or elastic-like (0.5:6) behavior across all $\dot{\gamma}$.

Intriguingly, $A_{F,i}$ decreases (and $\beta_{1,i}$ increases) with increasing $F_{R,i}$ for 0.5:0 and 0.5:6.4 whereas the opposite trend is seen for the 0.65:0 system. We can understand this phenomenon as arising from the varying degree to which DNA builds up or is compressed at the leading edge of the probe versus preferentially aligning with the strain path. Namely, reduced alignment and flow (smaller $A_{F,i}$). coupled to increased elastic storage (larger $F_{R,i}$), as seen for 0.5:0 and 0.5:6, indicates that it is polymer buildup at the strain edges, which increases the local entanglement density and propensity for threading, that, in turn, reduces the ability of the polymers to deform and move in response to the strain (reducing/increasing $A_{F,i}/\beta_{1,i}$). This phenomenon increases the polymer relaxation times, resulting in reduced dissipation and increased ability to elastically store force (increasing $F_{R,i}$). Conversely, the coupled increase of alignment and flow with elasticity for the 0.65:0 system is an indicator that increased shear-thinning and affine deformation/stretching (increasing/decreasing $A_{F,i}/\beta_{1,i}$) arise from increasingly strong entanglements and threading with increasing $\dot{\gamma}$ which increases the elasticity ($F_{R,i}$). The strong entanglements prevent polymer compression, as described above, by polymers being pulled in the direction of the strain from entanglements by leading polymers as well as opposite the strain from entanglements by trailing polymers. The result is increased entropic stretching and affine deformation, which, in turn, suppresses buildup at the strain edges.

**Conclusions**

Here, we present a robust experimental approach that couples macromolecular dynamics with local force response and stress propagation in polymeric fluids. OpTiDDM (Optical Tweezers integrating Differential Dynamic Microscopy) achieves this coupling by simultaneously imposing local nonlinear strains, measuring the forces the polymers exert to resist the strain, and visualizing the polymers surrounding the strain path. Our DDM analyses map the polymer deformation field as a function of distance from the strain path ($y$), correlation lengthscale ($\lambda = 2\pi/q_c$), strain rate ($\dot{\gamma}$) and network composition ($c_D:c_T$). We quantify the degree of polymer alignment with the strain path ($A_F$) and couple this alignment to the polymer relaxation dynamics (i.e., $\tau(q), \beta, \delta$).

We discover a surprising non-monotonic resonant response for entangled DNA blends, in which affine polymer alignment, superdiffusive motion, and elastic retention of induced force are maximized when $\dot{\gamma}$ is comparable to the entanglement rate $\upsilon_e$. Above or below this resonant rate, and for less entangled DNA, these effects are reduced. Incorporating microtubules into the DNA systems suppresses this resonance, as well as the $\dot{\gamma}$ dependence of polymer alignment and superdiffusivity, while at the same time substantially increasing the resistive force and elastic force retention. Notably, MTs also reduce the degree of subdiffusion (i.e., smaller $\beta$), despite the rigid scaffolding and caging they provide, while the most loosely connected system (0.5:0) exhibits the most pronounced subdiffusion.

We rationalize our results as arising from the varying propensity for polymers to buildup at the edges versus align and entropically stretch along the strain path, along with the varying spatiotemporal scales of entanglements and the degree to which polymers can dissipate stress via deformation and reorientation. These phenomena are likely dictated by the coupled effects of ring threading, entanglement fluctuations, and caging by rigid microtubules.

We acknowledge that the systems we chose to investigate to demonstrate the efficacy and applicability of OpTiDDM are quite complex, such that our interpretations are at times speculative and warrant future studies to test the hypothesized mechanisms. However, a primary goal of this study is to demonstrate how rich of a phase space of physical properties and relationships one can extract from OpTiDDM, and how well one can tease apart and couple together different observed phenomena in even the most complex of systems. We also acknowledge that the compositional differences between the systems we study are rather modest, which we purposefully chose to demonstrate the sensitivity of OpTiDDM to measure robust differences in mechanics and dynamics of similar systems.

While outside the scope of the current work, we hope that the approach and results we present will spur further experimental studies employing OpTiDDM to expand on the formulation space of the systems we examine. We also anticipate sparking new theoretical studies to develop predictive models for OpTiDDM metrics, and to describe the stress-dynamics phase space of complex polymeric blends and composites.

The precise and rich multi-scale characterization and coupling of mechanics and dynamics that OpTiDDM affords–along with its unique ability to map stress propagation and deformation fields resulting from programmable local strains–represents a major advance in the study of soft matter, complex fluids, polymer networks, viscoelastic materials and even biological cells. With its robust suite of capabilities, as well as its modular and adaptable design, we anticipate broad interdisciplinary use of OpTiDDM to elucidate non-trivial phenomena that dictate and are impacted by the propagation of local stresses through a system–critically important to commercial materials applications and cell mechanics alike.

**Methods**

**DNA**: Double-stranded, 115 kilobasepair (kbp) DNA, is prepared by replication of cloned bacterial artificial chromosomes (BACs) in *Escherichia coli*, followed by extraction, purification, concentration and resuspension in TE10 buffer (10 mM Tris-HCl (pH 8), 1 mM EDTA, 10 mM NaCl) using custom-designed protocols described and validated previously[85,86]. We use gel electrophoresis image analysis to quantify the mass concentration of the resulting DNA stock solutions $c_D$ and the corresponding mass fractions $f$ of relaxed circular (ring, $R$) and linear ($L$) topologies (no supercoiled constructs are present). Using Life Technologies E-Gel Imager and Gel Quant Express software we determine $c \simeq 0.8$ mg/ml, $f_L \simeq 0.5$ and $f_R \simeq 0.5$. We perform no enzymatic digestion after this point as we aimed to study ring-linear blends with $f_L \simeq f_R$ due to the intriguing emergent mechanical properties that have been previously reported in ring-linear blends of comparable fractions[35,42,45,75]. For experiments, we use two different dilutions of this stock: $c_{D,\downarrow} = 0.5$ mg/ml and $c_{D,\uparrow} = 0.65$ mg/ml.

The contour lengths of the ring and linear DNA are fixed at $L =$ 115 kbp (~38 μm), but due to their end-closure, rings have a ~$1.6x$ smaller radius of gyration $R_G$ than their linear counterparts[85]. Using the previously reported values of $R_{G,R} \simeq 541$ nm and $R_{G,L} \simeq 885$ nm for ring and linear 115 kbp DNA[85], we determine the polymer overlap concentration of the ring-linear DNA blend via $c_{RL}^* = (3/4\pi)(M/N_A)/(f_L R_{G,L}^3 + f_R R_{G,R}^3)$ where $M$ is the DNA molecular weight[1]. Our computed value of $c_{RL}^* \simeq 710$ μg/mL yields $c_{D,\downarrow} \simeq 7 c_{RL}^*$ and $c_{D,\uparrow} \simeq 9 c_{RL}^*$ for the two DNA concentrations we use in experiments.

To image DNA for DDM, we fluorescent-label a small fraction of the DNA molecules with MFP488 (Mirus) using the manufacturer-supplied *Label* IT Labeling Kit and corresponding protocols (Mirus). The excitation/emission spectrum for MFP488 is 501/523 nm and the dye molecule to DNA base pair ratio is 5:1.

**Microtubules (MT)**: Unlabeled porcine brain tubulin (T240) and rhodamine-labeled porcine brain tubulin (TL590M) are obtained from Cytoskeleton, Inc. 45.5 μM stock solutions of tubulin dimers, containing 9:1 unlabeled:labeled tubulin dimers in PEM100 buffer (100 mM PIPES (pH 6.8), 2 mM MgCl$_2$, 2mM EGTA) are flash-frozen in and stored at -80°C. As previously described[43], to form the DNA-MT composite, we add a molar concentration of $c_T = 6.4$ μM tubulin dimers to the 0.5 mg/ml DNA solution, followed by 2 mM GTP and 10 μM Taxol to polymerize the tubulin into microtubules and stabilize the MTs. The formed MTs are hollow rods with a diameter $D \simeq 25$ nm comprising 13 tubulin dimers per ring[87].

**Sample preparation**: We perform measurements on three different systems, which we denote by the ratio of the DNA mass concentration (in mg/ml) and tubulin molar concentration (in μM): $c_D : c_T = 0.5:0, 0.65:0$, and 0.5:6.4. We add 10 μg/ml MFP488-labeled DNA and a trace amount of polystyrene beads with radius $b =$ 2.25 μm (Polysciences, Inc.) to enable DDM and OpT measurements, respectively. Beads are coated with AlexaFluor594-BSA (ThermoFisher) to prevent DNA adsorption and enable fluorescence imaging. Note that beads and DNA are labeled with distinct excitation/emission dyes such that the beads do not interfere with DDM measurements of DNA tracers, and the fluorescence lifetimes of the DNA tracers are prolonged while moving to a new bead and location in the chamber. An oxygen scavenging system (45 μg/mL glucose, 43 μg/mL glucose oxidase, 7 μg/mL catalase, and 5 μg/mL β-mercaptoethanol) is added to inhibit photobleaching. Sample chambers (20×3×0.1 mm$^3$) are made with a microscope glass slide and coverslip, each coated with BSA to prevent adsorption of DNA, MTs and beads, and separated by two layers of double-sided tape. All samples are mixed slowly and thoroughly, using wide-bore pipette tips to prevent shearing of DNA, then introduced into sample chambers through capillary action and hermetically sealed with epoxy. For the DNA-MT composite ($c_D : c_T = 0.5:6.4$), tubulin dimers are added immediately before flowing into the chamber, and the sample chamber is incubated at 37°C for 2 hours, resulting in repeatable and reliable tubulin polymerization in the DNA solution[43].

**OpTiDDM instrumentation and protocol**: We use a custom-built optical trap formed from a 1064 nm Nd:YAG fiber laser (Manlight) focused with a 60× 1.4 NA objective (Olympus) and integrated into an Olympus IX71 epifluorescence microscope, as previously described[17]. The force exerted by the sample on a trapped bead is determined by measuring the laser beam deflections via a position sensing detector (PSD, First Sensor) at 20 kHz. The trap is calibrated for force measurement using the Stokes drag method[88]. To image the MPF488-labeled DNA and AlexaFluor594-labeled microspheres in the samples we use 490/525 nm and 530/575 nm excitation/emission filter cubes and an ORCA-Flash 4.0 LT+ CMOS camera (Hamamatsu). For DDM measurements, we use a piezoelectric actuator mirror (PI USA) to move the trap relative to the sample chamber while keeping the 600×900 square-pixel (130 nm/pixel) field-of-view (FOV) of the camera fixed and centered at the resting trap position. For force measurements we use a piezoelectric nanopositioning stage (Mad City Laboratories) to move the sample relative to the fixed trap.

As shown in Fig 1, the microrheological strain program we apply consists of repeatedly moving the trapped bead back and forth horizontally (along the $x$-axis) through a strain distance $s = 15$ μm at roughly logarithmically spaced strain rates of $\dot{\gamma} = 9.4, 19, 42, 90, 189$ s$^{-1}$ which correlate to speeds of $v = 10 - 200$ μm/s via the relation $\dot{\gamma} = 3v/\sqrt{2}b$ (9.4-189 s$^{-1}$)[89]. We pause between each 15 μm sweep for a fixed cessation time $\Delta t_R = 3$ s to allow the polymers to relax. We perform each oscillatory strain program for a total time of 50 s, and do not start a new sweep unless there is enough time to complete a full forward-backward cycle in the allotted 50 s time period. These criteria result in 6,7,8,8 and 9 complete cycles for the $\dot{\gamma} = 9.4, 19, 42, 90,$ and $189$ s$^{-1}$ strains, respectively. For each measurement ≥10 trials are conducted, each with a new microsphere in a new unperturbed location. Presented data is an average of all trials. During each 50 s strain, we capture a time-series of 78 μm ×117 μm images (centered on the strain path) of the labeled DNA in the sample at 60 fps, and record the PSD signal (i.e., laser deflections) at 20 kHz.

**Differential Dynamic Microscopy (DDM):** To determine how the strain-induced dynamics and structure of the polymeric fluids depend on the orthogonal distance $y$ from the strain path (Figs 1-5), we divide the FOV into 128×128 square-pixel (16.6 μm)$^2$ ROIs, centered horizontally at the midpoint of the $s = 15$ μm strain path and shifted along the $\pm y$ direction in 16-pixel increments, with the bottom edge of the first ROI at $y = 0$ (and its center at $y = 8$ μm) and the farthest ROI centered at $y = 28$ μm. We analyze 10 ROIs in each of the $+y$ and $-y$ directions and average the $+/-$ data for each $y$, as they exhibit statistically indistinguishable dynamics (as we expect).

We use custom-written scripts (Python) to perform DDM analysis as previously described[38]. In brief, DDM takes two-dimensional Fourier transforms of differences between images separated by a range of lag times $\Delta t$ to quantify how the degree of correlation decays with $\Delta t$ as a function of the wave vector $q$, which we quantify via a 3D image structure function $D(\vec{q}, \Delta t)$.

To determine the extent to which the DNA dynamics are preferentially aligned along the strain path ($x$-axis) (Fig 2), we adopt alignment factor analysis typically used to analyze scattering data produced by an aligned field[68,90]. We compute an alignment factor $A_F(q, \Delta t)$ with respect to the strain path from $D(\vec{q}, \Delta t)$ using:

$$A_F(q, \Delta t) = \frac{\int_0^{2\pi} D(q, \Delta t, \theta) \cos(2\theta) d\theta}{\int_0^{2\pi} D(q, \Delta t, \theta) d\theta} \qquad (1)$$

where $\theta$ is the angle relative to the $x$-axis[68]. To obtain a single $A_F$ value for each orthogonal distance $y$, we average over $\Delta t = 0.17$-1 s and $q = 1$-7 μm$^{-1}$, as there is no statistically significant dependence of $A_F(y)$ on these parameters. According to Eq. 1, $A_F$ can range between 0 (no alignment) and 1 (complete alignment), with increasing fractional values indicating more alignment.

To determine the type and rate of motion of the DNA, we radially average each $D(\vec{q}, \Delta t)$ to get a 1D image structure that can be described by the function $D(q, \Delta t) = A(q)[1 - f(q, \Delta t)] + B(q)$, where $f(q, \Delta t)$ is the intermediate scattering function (ISF), $A(q)$ is the amplitude, and $B(q)$ is the background. While radially averaging is only strictly valid for isotropic $D(\vec{q}, \Delta t)$ functions, which is admittedly a crude approximation in cases in which $A_F$ is large, we note that the largest $A_F$ value we measure is ~0.02, so the anisotropy is relatively weak even in the most extreme cases. To determine the type of motion and the corresponding rate, we model the ISF as a stretched exponential: $f(q, \Delta t) = e^{-(\Delta t/\tau(q))^\delta}$ where $\tau(q)$ is the decay time and $\delta$ the stretching exponent. Stretched ($\delta > 1$) and compressed ($\delta < 1$) exponentials, as opposed to simple exponentials ($\delta = 1$), have been shown to describe crowded, entangled[20,22,66,69,73] and active systems[70-72], respectively. The scaling of the $q$-dependent decay time $\tau(q) \sim q^{-\beta}$ describes the type of motion. $\beta = 2$ is indicative of normal Brownian diffusion whereas $\beta \to 1$ describes superdiffusive or ballistic dynamics and $\beta > 2$ indicates anamolous subdiffusion[10,21].

**Predicted length and time scales:** According to traditional reptation theory, the tube radius for linear polymers is computed using the relation $a_L = (24N_e/5)^{1/2} R_{G,L}$, where the number of entanglements per chain $N_e$ is given by $N_{e,L} = 4cRT/5MG_N^0$ where $G_N^0$ is the elastic plateau modulus[1]. Using our previously reported value of $G_N^0 \simeq 0.2$ Pa for linear 115 kbp DNA at $c = 1$ mg/ml[11], and the accepted scaling $G_N^0 \sim c^2$ for entangled linear polymers[1], we estimate $a_L$ for the $f_L \simeq 0.5$ fraction of linear DNA in the 0.5:0 and 0.65:0 blends (i.e., ~0.25 mg/ml and ~0.375 mg/ml) as $a_L \simeq 1.4$ μm and ~1.2 μm, respectively. To estimate the corresponding tube radius for the $f_R \simeq 0.5$ ring fraction, we use the pom-pom ring model prediction $a_R/a_L = (5N^{-0.4})^{1/2}$, where $N = 382$ is the number of Kuhn lengths[91], to compute $a_R \simeq 0.95$ μm and ~0.84 μm for $c_D = 0.5$ mg/ml and 0.65 mg/ml, respectively. Finally, to estimate an effective tube diameter $d_T = 2a$ for each $f_L \simeq f_R \simeq 0.5$ blend, we use the relation $d_T^{-3} = [(2a_L)^{-3} + (2a_R)^{-3}]$ that considers the density of each cubic tube diameter $d_T^3$ to arrive at values of $d_T \simeq 1.74$ μm and ~1.53 μm for the 0.5:0 and 0.65:0 blends, respectively[36]. For reference, if we assume that the blends are comprised entirely of linear chains we compute tube diameters of $d_{T,L} \simeq 1.98$ μm and ~1.73 μm for the 0.5:0 and 0.65:0 blends, respectively.

We can approximate the theoretical entanglement time $\tau_e$, or the time it takes for a linear entangled polymer to 'feel' its entanglement tube, based on the predicted expression for entangled linear polymers: $\tau_e \simeq (N_e R_G/D_0)^2$ where $D_0$ is the dilute limit diffusion coefficient[1], providing $\tau_{e,0.5} \simeq 15.7$ ms and $\tau_{e,0.65} \simeq 26.6$ ms for the 0.5:0 and 0.65:0 blends, respectively. Within the same tube theory framework, we can estimate the disengagement time $\tau_D$, i.e. the longest predicted relaxation time for entangled linear polymers, using $\tau_D \simeq (36R_G^4/\pi^2 a^2 D_0)$, providing $\tau_{D,0.5} \simeq 18.5$ s and $\tau_{D,0.65} \simeq 24.2$ s [1].

**Author Contributions:** RMRA conceived the project, guided the experiments, interpreted the data, and wrote the manuscript. KRP designed and performed the experiments, analyzed and interpreted the data, and wrote the manuscript. RC and PN helped prepare samples, perform experiments and analyze data. RJM developed analysis software and helped analyze data and write the manuscript.

**Competing Interest Statement:** The authors declare no competing interests.

**Acknowledgments:** This research was funded by grants from the Air Force Office of Scientific Research (AFOSR- FA9550-17-1-0249, AFOSR-FA9550-21-1-0361) awarded to RMRA.

**OpTiDDM (Optical Tweezers integrating Differential Dynamic Microscopy) maps the spatiotemporal propagation of nonlinear stresses in polymer blends and composites**

Karthik R. Peddireddy, Ryan Clairmont, Philip Neill, Ryan McGorty, and Rae M. Robertson-Anderson*

*Department of Physics and Biophysics, University of San Diego, 5998 Alcala Park, San Diego, CA 92110, United States*

**Supplemental Information**

**Figure S1.** Confocal micrographs of 0.5:6.4 DNA-MT composite.

**Figure S2.** OpTiDDM measurements in water.

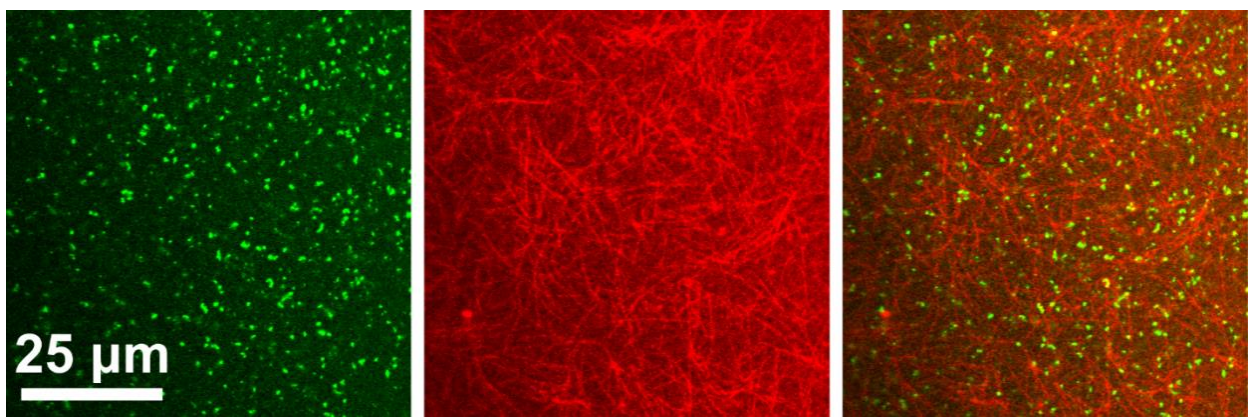

**Figure S1. Confocal micrographs of 0.5:6.4 DNA-MT composite.** Confocal micrographs (71 µm × 71 µm) of a DNA-MT composite with 6.4 µM rhodamine-labeled tubulin polymerized into microtubules in a solution of 0.5 mg/ml DNA (comprising a 1:1 ratio of ring and linear topologies). The green (left) and red (middle) images show MFP488-labeled DNA and rhodamine-labeled microtubules in the 0.5:6.4 composite, and the rightmost figure is the composite of the two channels. The scale bar applies to all images. The 512×512 square-pixel images (1 pixel = 0.14 µm) are acquired using a Nikon A1R laser scanning confocal microscope with a 60× 1.4 NA objective. 10% of tubulin dimers comprising microtubules are rhodamine-labeled to enable imaging using a 561 nm laser with 561 nm excitation and 595 nm emission filters. MFP488-labeled DNA, comprising 10% of the total DNA in solution, are imaged using a 488 nm laser with 488 nm excitation and 525 nm emission filters.

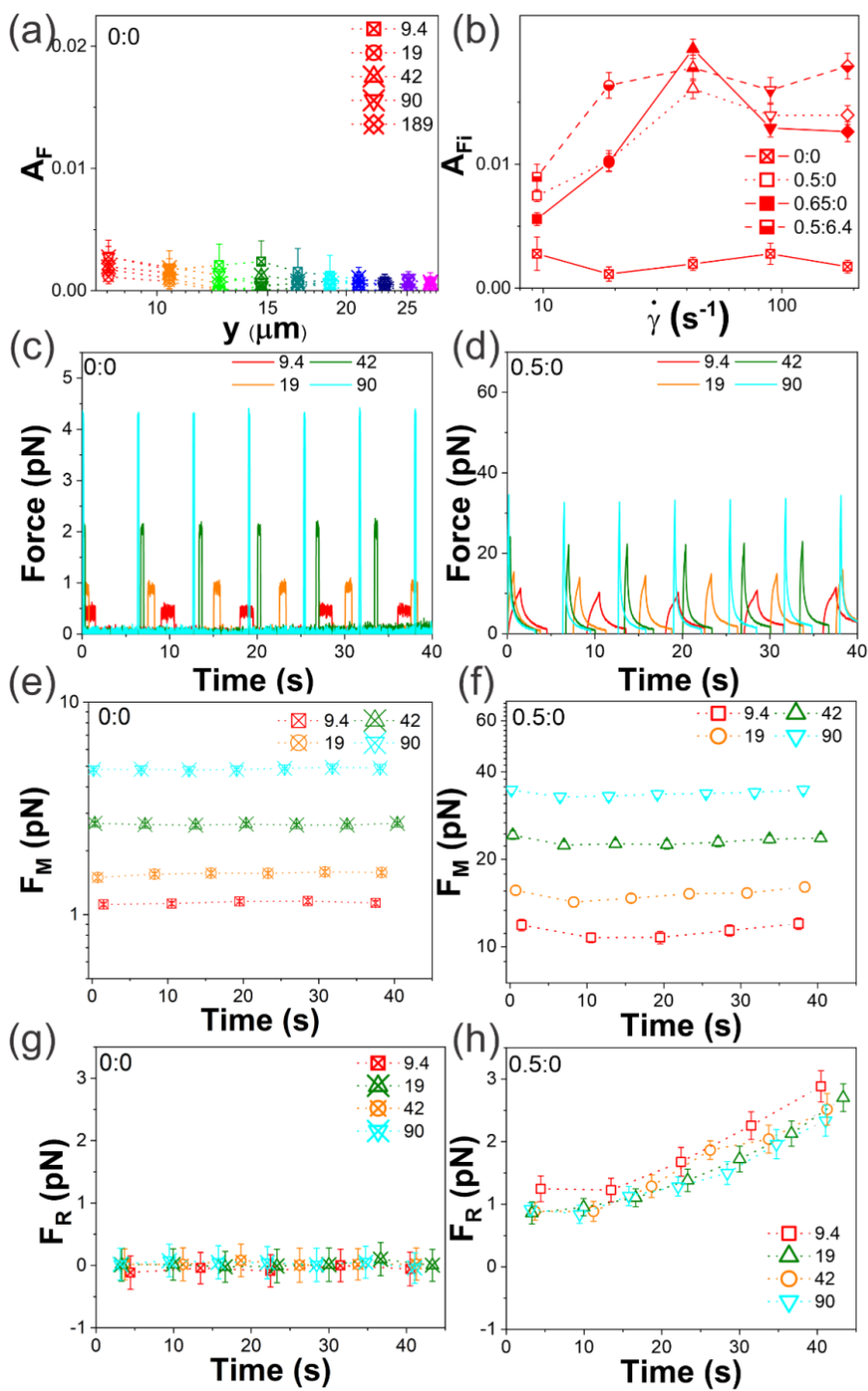

**Figure S2. OpTiDDM control measurements in water.** (a) $A_F(y)$ for water predictably shows no significant alignment in the direction of strain path for all $\dot{\gamma}$ and $y$ values. (b) $A_F$ versus $\dot{\gamma}$ for the closest ROI ($y =8$ μm) shows that $A_F$ in water is negligibly small and independent of $\dot{\gamma}$ compared to the polymer systems we study here, $c_D$: $c_T = 0.5{:}0$, $0.65{:}0$ and $0.5{:}6.4$ (c-d) Sample force traces for cyclic strains in water (c) and the 0.5:0 blend (d) performed at strain rates listed in the legend in s$^{-1}$. (e-h) Maximum force $F_M$ (e,f) and residual force $F_R$ (g,h) measured in water (e,g) compared to 0.5:0 (f,h) for strain rates listed in the legends in s$^{-1}$.